\documentclass[onecolumn,showpacs,preprintnumbers]{revtex4}
\usepackage{graphicx}
\usepackage{dcolumn}
\usepackage{bm}
\usepackage{epsfig}
\usepackage{amsmath,amssymb}
\usepackage{caption2}
\usepackage{subfigure}
\usepackage{mathrsfs}

\begin{document}
\title{Analysis of the strong coupling constant $G_{D_{s}^{*}D_{s}\phi}$ and the decay width
of $D_{s}^{*}\rightarrow D_{s}\gamma$ with QCD sum rules }
\author{Guo-Liang Yu$^{1}$}
\email{yuguoliang2011@163.com}
\author{Zhen-Yu Li$^{2}$}
\author{Zhi-Gang Wang$^{1}$}
\email{zgwang@aliyun.com}

\affiliation{$^1$ Department of Mathematics and Physics, North China
Electric power university, Baoding 071003, People's Republic of
China\\$^2$ School of Physics and Electronic Science, Guizhou Normal
College, Guiyang 550018, People's Republic of China}
\date{\today }

\begin{abstract}
In this article, we calculate the form factors and the coupling
constant of the vertex $D_{s}^{*}D_{s}\phi$ using the three-point
QCD sum rules. We consider the contributions of the vacuum
condensates up to dimension $7$ in the operator product
expansion(OPE). And all possible off-shell cases are considered,
$\phi$, $D_{s}$ and $D_{s}^{*}$, resulting in three different form
factors. Then we fit the form factors into analytical functions and
extrapolate them into time-like regions, which giving the coupling
constant for the process. Our analysis indicates that the coupling
constant for this vertex is $G_{Ds*Ds\phi}=4.12\pm0.70 GeV^{-1}$.
The results of this work are very useful in the other
phenomenological analysis. As an application, we calculate the
coupling constant for the decay channel $D_{s}^{*}\rightarrow
D_{s}\gamma$ and analyze the width of this decay with the assumption
of the vector meson dominance of the intermediate $\phi(1020)$. Our
final result about the decay width of this decay channel is
$\Gamma=0.59\pm0.15keV$.
\end{abstract}


\maketitle

\begin{large}
\textbf{1 Introduction}
\end{large}

In relativistic heavy ion collisions $J/\psi$ suppression has been
recognized as an important tool to identify the possible phase
transition to quark-gluon plasma\cite{Mats86}. The dissociation of
$J/\psi$ in the quark-gluon plasma due to color screening can lead
to a reduction of its production. People usally explained this
phenomenon as a process of the $J/\psi$ absorption by $\pi$, $\rho$
or $\phi$ mesons in a meson-exchange model\cite{Mati98}. And we can
calculate the the absorption cross sections based on the
interractions among the quarkonia and mesons, where the hadronic
coupling constants are basic input parameters. A detailed knowledge
of the hadronic coupling constants is of great importance in
understanding the effects of heavy quarkonium absorptions in
hadronic matter. Besides, the hadronic coupling constants about the
heavy-light mesons can also help us understanding the final-state
interacions in the heavy quarkonium decays\cite{Casa97}.
Furthermore, some exotic mesons have been detected in recent years
\cite{Aal09}, which are beyond the usual quark-model description as
$q\overline{q}$ pairs. And people interpret them as quark-gluon
hybrids ($q\overline{q}g$), tetraquark states
($q\overline{q}q\overline{q}$), molecular states of two ordinary
mesons, glueballs, states with exotic quantum numbers and many
others\cite{Mah09}. The form factors and coupling constants play an
important role in understanding the nature of these exotic mesons.

However, the strong coupling constant used in the above questions
can not be explained by perturbative theories, because the associate
interactions lie in the low energy region. It is fortunate that the
QCDSR approach can help us to solve the difficulty. The QCDSR is one
of the most powerful non-perturbative methods, which is also
independent of model parameters. In recent years, numerous research
articles have been reported about the precise determination of the
strong form factors and coupling constants via QCDSR, light-cone
QCDSR or lattice calculation\cite{Aydin04,Aziz11,Doi04}. And many
strong coupling constants have been determined by different groups,
for example, $D^{*}D_{s}K$, $D{s}^{*}DK$, $B_{c}^{*}B_{c}\Upsilon$,
$B_{c}^{*}B_{c}\psi$, $B_{s}^{*}BK$, $J/\psi D_{s}^{*}D_{s}$,
$J/\psi D_{s}D_{s}$, $J/\psi D_{s}^{*}D_{s}^{*}$,
$D_{s}^{*}D_{s}\eta'$\cite{Aydin04,ZGW14,Kohs14}. In this work, we
use the QCDSR formalism to obtain the coupling constant of the meson
vertice $ D_{s}^{*}D_{s}\phi$, where the contributions of the vacuum
condensates up to dimension $7$ in the OPE are considered. The
results of this work are very useful in these phenomenological
analysis mentioned above.

It is indicated by the BaBar collaboration that
$\Gamma(D_{s}^{*})<1.9MeV$ and $\frac{\Gamma(D_{s}^{*}\rightarrow
D_{s}\gamma)}{\Gamma_{Total}}\approx0.94$\cite{Aub052}. However, the
exact value of the decay width have yet not been determined. A more
exact result can help us understanding the nature of the meson and
testing the validity of the theoretical model. As an application, we
also give an analysis about the decay $D_{s}^{*}\rightarrow
D_{s}\gamma$ in the end of this paper, where the electromagnetic
coupling constant $G_{D_{s}^{*}D_{s}\gamma}$ will be used. This
coupling constant can be easily obtained, when we set $Q^{2}=0$ in
the analytical function of coupling constant
$G_{D^{*}_{s}D_{s}\phi}(Q^2)$ in Sec.III.

The outline of this paper is as follows. In Sec.II, we study the $
D_{s}^{*}D_{s}\phi$ vertices using the three-point QCDSR. In order
to reduce the uncertainties of the result, we calculate the
three-point correlation functions: one with the vector meson $\phi$
off-shell, another with the pseudoscalar meson $D_{s}$ off-shell,
and a third one with the vector meson $D^{*}_{s}$ off-shell. Besides
of the perturbative contribution, we also consider the contribution
of $\langle q\overline{q}\rangle$, $\left\langle \overline{q}g\sigma
.Gq\right\rangle$, $\langle g^2G^2\rangle$, $\langle f^3G^3\rangle$,
$\langle q\overline{q} \rangle^2$ and $\langle q\overline{q}\rangle
\langle GG\rangle$ at OPE side. In Sec. III, we present the
numerical results and discussions, and Sec IV is reserved for our
conclusions.

\begin{large}
\textbf{2 QCD sum rules for the $ D_{s}^{*}D_{s}\phi$}
\end{large}

In this work, the $D_{s}^{*}D_{s}\phi$ is a
vector-pseudoscalar-vector($VPV$) vertex. With each meson off-shell,
we write down the three-point correlation functions:

\begin{eqnarray}
&&\Pi_{\mu\nu}^{\phi}(p,p^{\prime})=i^{2}\int d^{4}%
xd^{4}ye^{ip^{\prime}.x+i(p-p^{\prime}).y}\left\langle
0|T\{J_{5}(x)j_{\mu }(y)J_{\nu}^{\dagger}(0)\}0|\right\rangle \\
&&\Pi_{\mu\nu}^{D_{s}}(p,p^{\prime})=i^{2}\int
d^{4}xd^{4}ye^{ip^{\prime }.x+i(p-p^{\prime}).y}\left\langle
0|T\{J_{\nu}(x)J_{5}(y)j_{\mu}^{\dagger }(0)\}0|\right\rangle \\
&&\Pi_{\mu\nu}^{D^{*}_{s}}(p,p^{\prime})=i^{2}\int d^{4}%
xd^{4}ye^{ip^{\prime}.x+i(p-p^{\prime}).y}\left\langle
0|T\{j_{\mu}(x)J_{\nu}(y)J_{5}^{\dagger }(0)\}0|\right\rangle
\end{eqnarray}
where $T$ is the time ordered product and $J_{\nu}^{\dagger}(x)$,
$J_{5}(x)$ and $j_{\mu}(x)$ are the interpolating currents of the
mesons $D_{s}^{*}$, $D_{s}$ and $\phi$ respectively:

\begin{eqnarray}
&&J_{\nu}^{\dagger}(x)=\bar{s}(x)\gamma_{\nu}c(x) \\
&&J_{5}(x)=\bar{c}(x)i\gamma_{5}s(x) \\
&&j_{\mu}(x)=\bar{s}(x)\gamma_{\mu}s(x)
\end{eqnarray}
According to the QCDSR, these correlation functions can be
calculated in two different ways: using hadron degrees of freedom,
called the phenomenological side, or using quark degrees of freedom,
called the OPE side. In the following we will obtain the sum rule
according to above formulations.

\begin{large}
\textbf{2.1 The phenomenological side}
\end{large}

We insert a complete set of intermediate hadronic states with the
same quantum numbers as the current operators
$J_{\nu}^{\dagger}(x)$, $J_{5}(x)$ and $j_{\mu}(x)$ into the
correlation functions $\Pi_{\mu\nu}^{\phi}(p,p^{\prime})$,
$\Pi_{\mu\nu}^{D_{s}}(p,p^{\prime})$ and
$\Pi_{\mu\nu}^{D^{*}_{s}}(p,p^{\prime})$ to obtain the
phenomenological representations. After isolating the ground-state
contributions, we get the following functions for the mesons $\phi$,
$D_{s}$ and $D^{*}_{s}$ off-shell cases.
\begin{eqnarray}
&&\Pi_{\mu\nu}^{phen(\phi)}=\frac{-CG_{D_{s}^{\ast}D_{s}\phi}^{(\phi)}%
(q^{2})p^{\alpha}p^{\prime\beta}\varepsilon_{\mu\nu\alpha\beta}}%
{(p^{2}+m_{D_{s}^{\ast}}^{2})(q^{2}+m_{\phi}^{2})(p^{\prime2}+m_{D_{s}}^{2})}+h.r.
\\
&&\Pi_{\mu\nu}^{phen(D_{s})}=\frac{-CG_{D_{s}^{\ast}D_{s}\phi}^{(D_{s})}%
(q^{2})p^{\alpha}p^{\prime\beta}\varepsilon_{\mu\nu\alpha\beta}}%
{(p^{2}+m_{\phi}^{2})(q^{2}+m_{D_{s}}^{2})(p^{\prime2}+m_{D_{s}^{\ast}}^{2})}+h.r. \\
&&\Pi_{\mu\nu}^{phen(D_{s}^{\ast})}=\frac{-CG_{D_{s}^{\ast}D_{s}\phi}%
^{(D_{s}^{\ast})}(q^{2})p^{\alpha}p^{\prime\beta}\varepsilon_{\mu\nu
\alpha\beta}}{(p^{2}+m_{D_{s}}^{2})(q^{2}+m_{D_{s}^{\ast}}^{2})(p^{\prime
2}+m_{\phi}^{2})}+h.r.
\end{eqnarray}
where
$C=\frac{f_{D_{s}}m_{D_{s}}^{2}f_{D_{s}^{\ast}}m_{D_{s}^{\ast}%
}f_{\phi}m_{\phi}}{(m_{s}+m_{c})}$ and $h.r.$ stand for the
contributions of higher resonances and continuum states of each
meaon. And in the derivation, we have used the following effective
Lagrangian $\pounds$ and definitions for the decay constants
$f_{D_{s}^{\ast}}$, $f_{D_{s}}$ and $f_{\phi}$:

\begin{eqnarray}
&&\pounds
=G_{D_{s}^{\ast}D_{s}\phi}\varepsilon_{\alpha\beta\lambda\tau
}(\partial^{\alpha}D_{s}^{\ast +
\beta}D_{s}^{-}\partial^{\lambda}+\partial^{\alpha}D_{s}^{\ast -
\beta}D_{s}^{+}\partial^{\lambda})\phi^{\tau} \\
&&\left\langle 0|J_{\nu}(0)|D_{s}^{\ast}(p)\right\rangle =f_{D_{s}^{\ast}%
}m_{D_{s}^{\ast}}\zeta_{\mu}\\
&&\left\langle 0|J_{5}(0)|D_{s}(p^{\prime})\right\rangle =f_{D_{s}}m_{D_{s}%
}^{2}/(m_{s}+m_{c})\\
&&\left\langle  0|j_{\mu}(0)|\phi(q)\right\rangle
=f_{\phi}m_{\phi}\xi_{\mu}
\end{eqnarray}
where $\zeta_{\mu}$ and $\xi_{\mu}$ are the polarization vectors.
From Eqs.(7)$\sim$(9), we can see that there is only one tensor
structure to work within the formalism of the QCDSR.

\begin{large}
\textbf{2.2 The OPE side}
\end{large}

Now, we briefly outline the operator product expansion for the
correlation functions Eqs.(1)$\sim$(3) Firstly, we contract the
quark fields in the correlation functions with Wich's theorem.
\begin{eqnarray}
&&\Pi _{\mu \nu }^{(\phi) }=-i^{2}\int d^{4}xd^{4}ye^{ip^{\prime
}x+i(p-p^{\prime })y}tr\{i\gamma _{5}s^{mn}(x-y)\gamma _{\mu
}s^{nk}(y-0)\gamma _{\nu }c^{km}(0-x)\} \\
&&\Pi _{\mu \nu }^{(D_{s})}=-i^{2}\int d^{4}xd^{4}ye^{ip^{\prime
}x+i(p-p^{\prime })y}tr\{\gamma _{\nu }c^{mn}(x-y)i\gamma
_{5}s^{nk}(y-0)\gamma _{\mu }s^{km}(0-x)\} \\
&&\Pi _{\mu \nu }^{(D_{s}^{\ast })}=-i^{2}\int
d^{4}xd^{4}ye^{ip^{\prime }x+i(p-p^{\prime })y}tr\{\gamma _{\mu
}s^{mn}(x-y)\gamma _{\nu }c^{nk}(y-0)i\gamma _{5}s^{km}(0-x)\}
\end{eqnarray}
Then, we replace the $c$ and $s$ quark propagators $c^{ij}(x)$ and
$s^{ij}(x)$ with the corresponding full propagators\cite{Wang14},

\begin{eqnarray}
\notag\
S_{ij}(x)&&=\frac{i\delta _{ij}x \!\!\!/}{2\pi ^{2}x^{4}}-\frac{\delta _{ij}m_{s}}{%
4\pi ^{2}x^{4}}-\frac{\delta _{ij}\left\langle \overline{s}s\right\rangle }{%
12}+\frac{i\delta _{ij}x\!\!\!/ m_{s}\left\langle \overline{s}s\right\rangle }{48}-%
\frac{\delta _{ij}x^{2}\left\langle \overline{s}g_{s}\sigma
Gs\right\rangle }{192}+\frac{i\delta _{ij}x^{2}x\!\!\!/
m_{s}\left\langle \overline{s}g_{s}\sigma Gs\right\rangle }{1152}\\
\notag\ &&-\frac{ig_{s}G_{\alpha \beta }^{a}t_{ij}^{a}(x\!\!\!/
\sigma ^{\alpha \beta }+\sigma ^{\alpha \beta }x\!\!\!/)}{32\pi
^{2}x^{2}}-\frac{i\delta
_{ij}x^{2}x\!\!\!/g_{s}^{2}\left\langle \overline{s}s\right\rangle ^{2}}{7776}-%
\frac{\delta _{ij}x^{4}\left\langle \overline{s}s\right\rangle
\left\langle g_{s}^{2}GG\right\rangle }{27648}-\frac{\left\langle
\overline{s}_{j}\sigma ^{\mu \nu }s_{i}\right\rangle \sigma _{\mu
\nu }}{8}\\  && -\frac{\left\langle \overline{s}_{j}\gamma ^{\mu
}s_{i}\right\rangle \gamma _{\mu }}{4}+\cdot\cdot\cdot,
\end{eqnarray}

\begin{eqnarray}
\notag\
C_{ij}(x)&&=\frac{i}{(2\pi )^{4}}\int d^{4}ke^{-ik.x}\bigg \{\frac{\delta _{ij}}{%
k-m_{c}}-\frac{g_{s}G_{\alpha \beta }^{n}t_{ij}^{n}}{4}\frac{\sigma
^{\alpha\beta }(k\!\!\!/+m_{c})+(k\!\!\!/+m_{c})\sigma ^{\alpha
\beta }}{(k^{2}-m_{c}^{2})^{2}}\\
& &+\frac{g_{s}D_{\alpha }G_{\beta \lambda
}^{n}t_{ij}^{n}(f^{\lambda \beta \alpha }+f^{\lambda \alpha \beta
})}{3(k^{2}-m_{c}^{2})^{4}}
-\frac{g_{s}^{2}(t^{a}t^{b})_{ij}G_{\alpha \beta }^{a}G_{\mu \nu
}^{b}(f^{\alpha \beta \mu \nu }+f^{\alpha \mu \beta \nu }+f^{\alpha
\mu \nu \beta })}{4(k^{2}-m_{c}^{2})^{5}}+\cdot\cdot\cdot \bigg \},
\end{eqnarray}
\begin{eqnarray}
&&f^{\lambda \alpha \beta }=(k\!\!\!/+m_{c})\gamma ^{\lambda
}(k\!\!\!/+m_{c})\gamma ^{\alpha }(k\!\!\!/+m_{c})\gamma ^{\beta
}(k\!\!\!/+m_{c}) \\
&&f^{\alpha \beta \mu \nu }=(k\!\!\!/+m_{c})\gamma ^{\alpha
}(k\!\!\!/+m_{c})\gamma ^{\beta }(k\!\!\!/+m_{c})\gamma ^{\mu
}(k\!\!\!/+m_{c})\gamma ^{\nu }(k\!\!\!/+m_{c})
\end{eqnarray}
where $\langle g^{2}_{s}GG \rangle=\langle
g^{2}_{s}G^{n}_{\alpha\beta}G^{n\alpha\beta} \rangle$,
$t^{n}=\frac{\lambda^{n}}{2}$, the $\lambda^{n}$ is the Gell-Mann
matrix, $D_{\alpha}=\partial_{\alpha}-ig_{s}G^{n}_{\alpha}t^{n}$,
and the $i,j$ are color indices. Then we compute the integrals both
in the coordinate and momentum spaces, and obtain the correlation
functions. Finally, the correlation functions can be divided into
two parts:

\begin{eqnarray}
\Pi_{\mu\nu}^{OPE(M)}=\Pi_{\mu\nu}^{pert(M)}+\Pi_{\mu\nu}^{non-pert(M)}
\end{eqnarray}
where $M$ is the off-shell meson($M=\phi,D_{s},D^{*}_{s}$). Using
dispersion relations, the perturbative term for a given meson $M$
off-shell can be written in the following form:
\begin{eqnarray}
\Pi_{\mu\nu}^{pert(M)}(p,p^{\prime})=-\frac{1}{4\pi^{2}}\int_{0}^{\infty}%
\int_{0}^{\infty}\frac{\rho_{\mu\nu}^{pert(M)}(s,u,q^{2})}{(s-p^{2}%
)(u-p^{\prime2})}dsdu
\end{eqnarray}
and the quantities $s=p^{2}$, $u=p'^2$ and $q=p-p'$. We put all
quark lines on mass shell using Cutkosky's rules(Fig.1 ($a$) and
($b$)) and obtain the spectral density
$\rho_{\mu\nu}^{pert(M)}(s,u,q^{2})$

\begin{figure}[htp]
\begin{center}
\subfigure[]{\includegraphics[height=4.0cm,width=5.4cm]{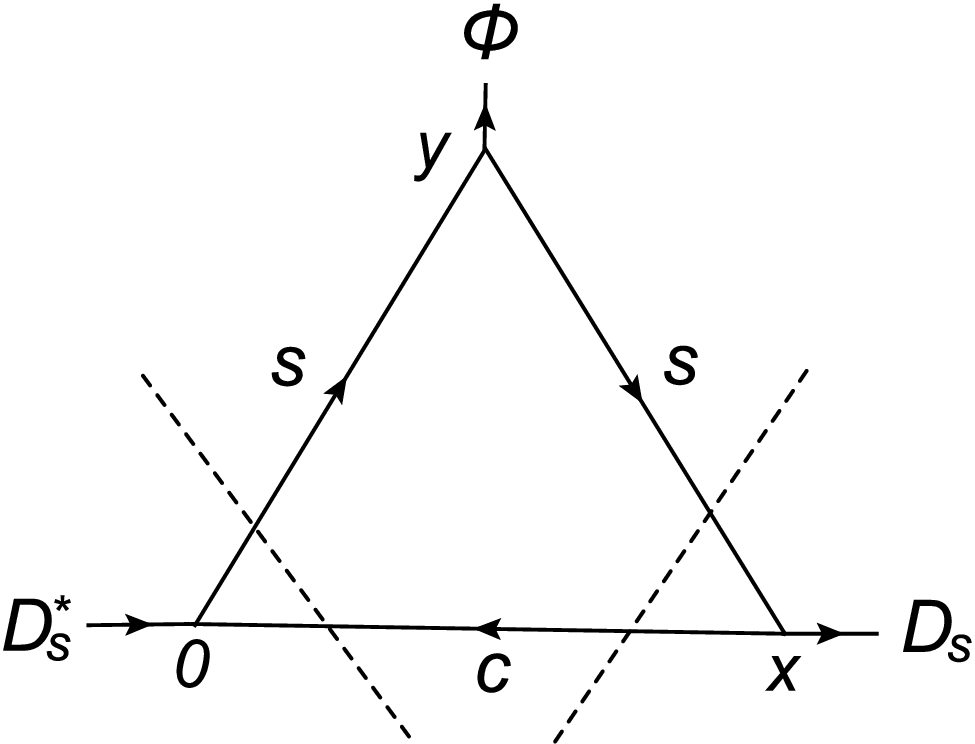}}
\subfigure[]{\includegraphics[height=4.1cm,width=5.6cm]{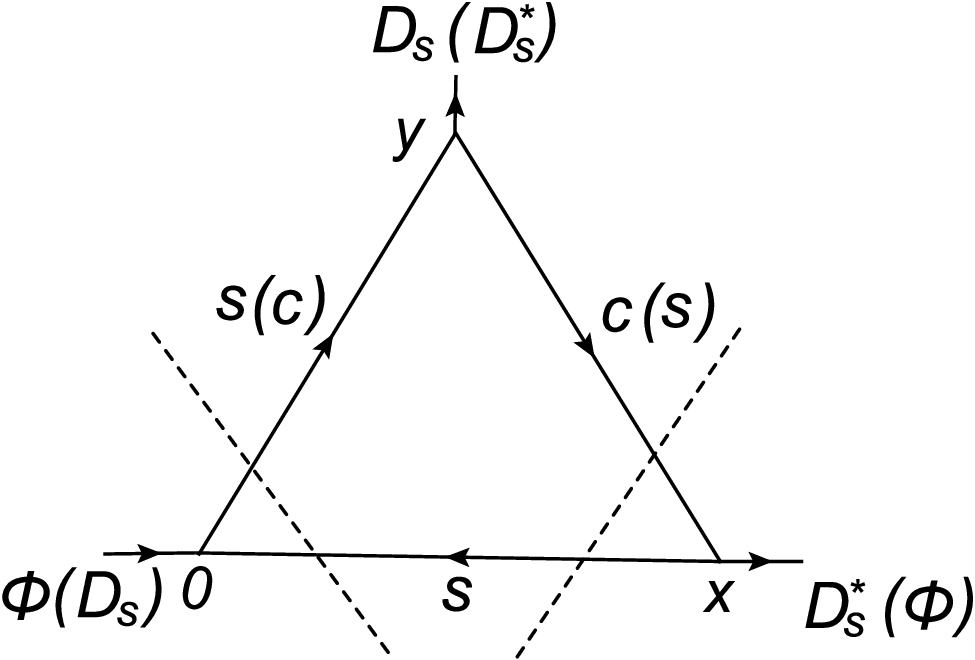}}
\end{center}
\caption{The perturbative contributions for $\phi$, $D_{s}$ and
$D_{s}^{*}$ off-shell. The Dashed lines denote the Cutkosky cuts.}
\end{figure}
\begin{eqnarray}
\rho_{\mu\nu}^{pert(\phi)}(s,u,q^{2})&=&-\frac{3}{\sqrt{\lambda}}\left[(m_{c}%
-m_{s})\frac{q^{2}(s+u-q^{2}+2m_{s}^{2}-2m_{c}^{2})}{\lambda(s,u,q^{2})}%
-m_{s}\right] \varepsilon_{\mu\nu\alpha\beta}p^{\alpha}p^{\prime\beta} \\
\notag\
\rho_{\mu\nu}^{pert(D_{s})}(s,u,q^{2})&=&-\frac{3}{\sqrt{\lambda}}\left[(m_{c}%
-m_{s})\frac{(u-q^{2})(s+u-q^{2})-2s(u+m_{c}^{2}-m_{s}^{2})}{\lambda
(s,u,q^{2})}-m_{c}\right]\varepsilon_{\mu\nu\alpha\beta}p^{\alpha}p^{\prime\beta}\\
\\ \notag
\rho_{\mu\nu}^{pert(D_{s}^{\ast})}(s,u,q^{2})&=&-\frac{3}{\sqrt{\lambda}%
}\left[(m_{c}-m_{s})\frac{u(s+u-q^{2})-2u(m_{c}^{2}-m_{s}^{2}+u-q^{2})}%
{\lambda(s,u,q^{2})}-m_{s}\right]\varepsilon_{\mu\nu\alpha\beta}p^{\alpha}%
p^{\prime\beta}\\
\end{eqnarray}
where $\lambda(s,u,q^{2}=(s+u-q^{2}))^{2}-4su$. As to the
non-perturbative contributions, we take into account the
contribution of $\langle s\overline{s}\rangle$, $\left\langle
\overline{s}g\sigma .Gs\right\rangle$, $\langle g^2G^2\rangle$,
$\langle f^3G^3\rangle$, $\langle s\overline{s} \rangle^2$ and
$\langle s\overline{s}\rangle \langle GG\rangle$, which are showed
explicitly in Figs 2 and 3. It should be noticed that as the
consequence of the use of the double Borel transform, the $\phi$
off-shell case has only the contributions of $\langle g^2G^2\rangle$
and $\langle f^3G^3\rangle$(Fig.2). Full expressions for these
contributions of Figs 2 and 3 for $\phi$, $D_{s}$ and $D^{*}_{s}$
off-shell cases can be found in Appendix A,B and C, where the
following representations will be used:

\begin{eqnarray}
\notag\
N_{m_{1}m_{2}m_{3}}^{abc}&=&\frac{(-1)^{a+b+c}\pi
^{2}i}{\Gamma (a)\Gamma (b)\Gamma
(c)(M_{1}^{2})^{b}(M_{2}^{2})^{c}(M^{2})^{a-2}}\int_{0}^{\infty
}d\tau (\tau +1)^{a+b+c-4}\tau ^{1-b-c} \\
&&\exp\left\{-\frac{1}{\tau}\frac{Q^{2}}{M_{1}^{2}+M_{2}^{2}}-\frac{(\tau
+1)m_{1}^{2}}{M^{2}}-\frac{(\tau +1)m_{2}^{2}}{\tau
M_{1}^{2}}-\frac{(\tau +1)m_{3}^{2}}{\tau M_{2}^{2}}\right\}
\end{eqnarray}
\begin{eqnarray}
\notag\
 I_{m_{1}m_{2}m_{3}}^{abc}&=&\frac{(-1)^{a+b+c}\pi
^{2}i}{\Gamma (a)\Gamma (b)\Gamma
(c)(M_{1}^{2})^{b}(M_{2}^{2})^{c+1}(M^{2})^{a-3}}\int_{0}^{\infty
}d\tau (\tau +1)^{a+b+c-5}\tau ^{1-b-c}\\
&&\exp \left\{-\frac{1}{\tau
}\frac{Q^{2}}{M_{1}^{2}+M_{2}^{2}}-\frac{(\tau
+1)m_{1}^{2}}{M^{2}}-\frac{(\tau +1)m_{2}^{2}}{\tau
M_{1}^{2}}-\frac{(\tau +1)m_{3}^{2}}{\tau M_{2}^{2}}\right\}
\end{eqnarray}
\begin{eqnarray}
\notag\
\widetilde{I}_{m_{1}m_{2}m_{3}}^{abc}&=&\frac{(-1)^{a+b+c}\pi
^{2}i}{\Gamma
(a)\Gamma (b)\Gamma (c)(M_{1}^{2})^{b+1}(M_{2}^{2})^{c}(M^{2})^{a-3}}%
\int_{0}^{\infty }d\tau (\tau +1)^{a+b+c-5}\tau ^{1-b-c}\\
&&\exp\left\{-\frac{1}{\tau
}\frac{Q^{2}}{M_{1}^{2}+M_{2}^{2}}-\frac{(\tau
+1)m_{1}^{2}}{M^{2}}-\frac{(\tau +1)m_{2}^{2}}{\tau
M_{1}^{2}}-\frac{(\tau +1)m_{3}^{2}}{\tau M_{2}^{2}}\right\}
\end{eqnarray}

\begin{figure}[htp]
\begin{center}
\subfigure[]{\includegraphics[height=2.0cm,width=2.6cm]{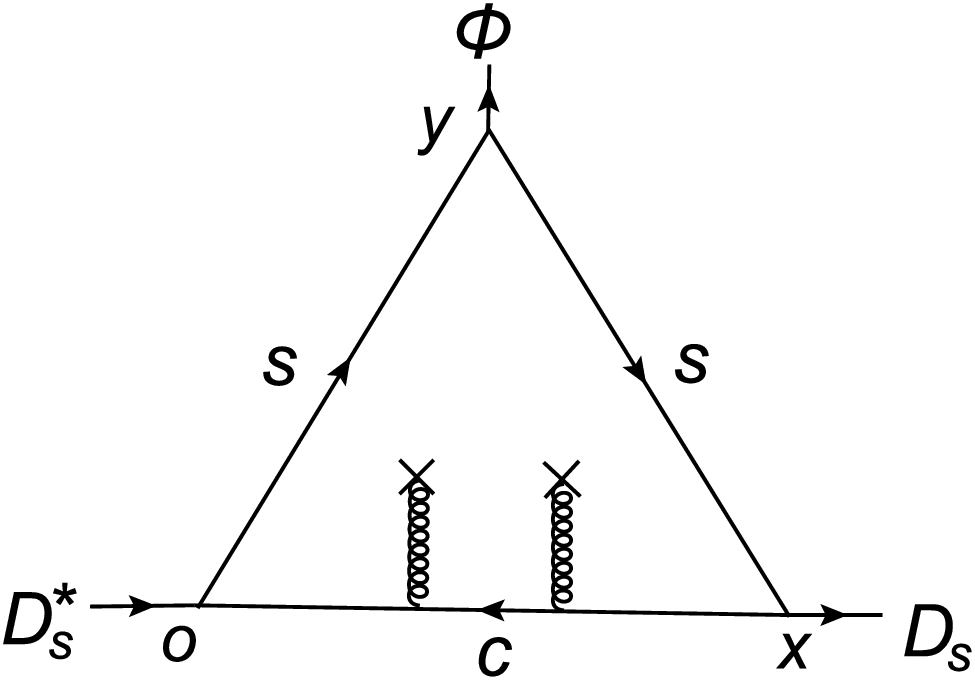}}
\subfigure[]{\includegraphics[height=2.0cm,width=2.6cm]{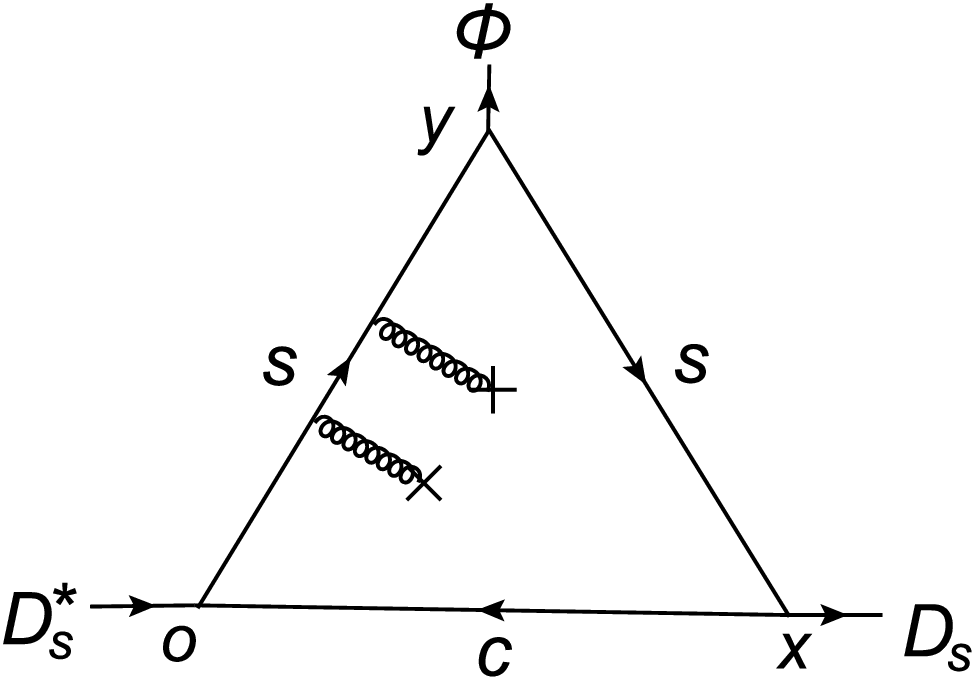}}
\subfigure[]{\includegraphics[height=2.0cm,width=2.6cm]{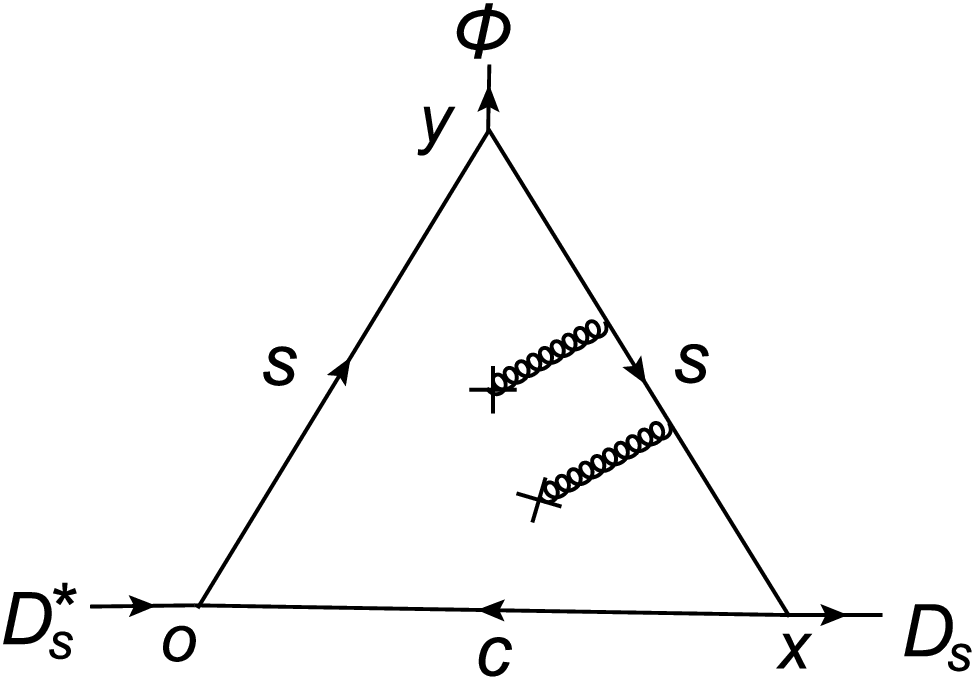}}
\subfigure[]{\includegraphics[height=2.0cm,width=2.6cm]{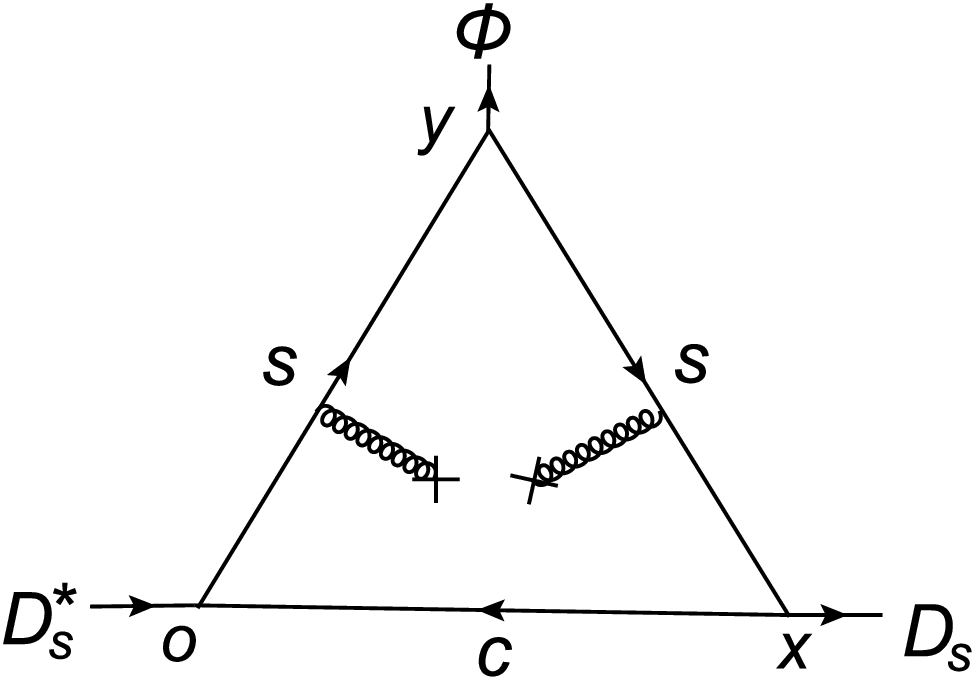}}

\subfigure[]{\includegraphics[height=2.0cm,width=2.6cm]{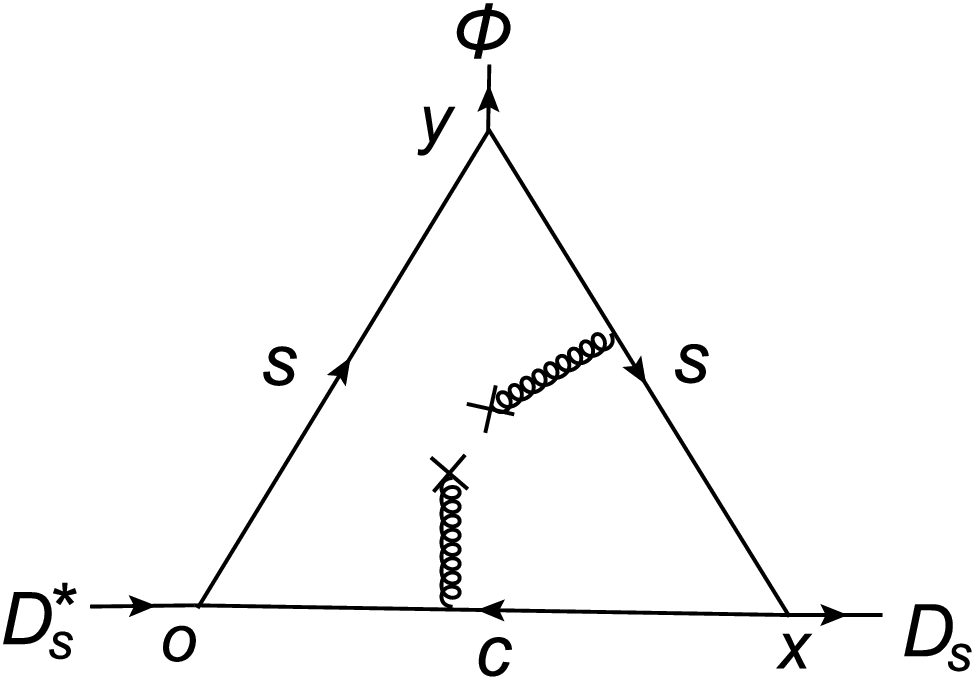}}
\subfigure[]{\includegraphics[height=2.0cm,width=2.6cm]{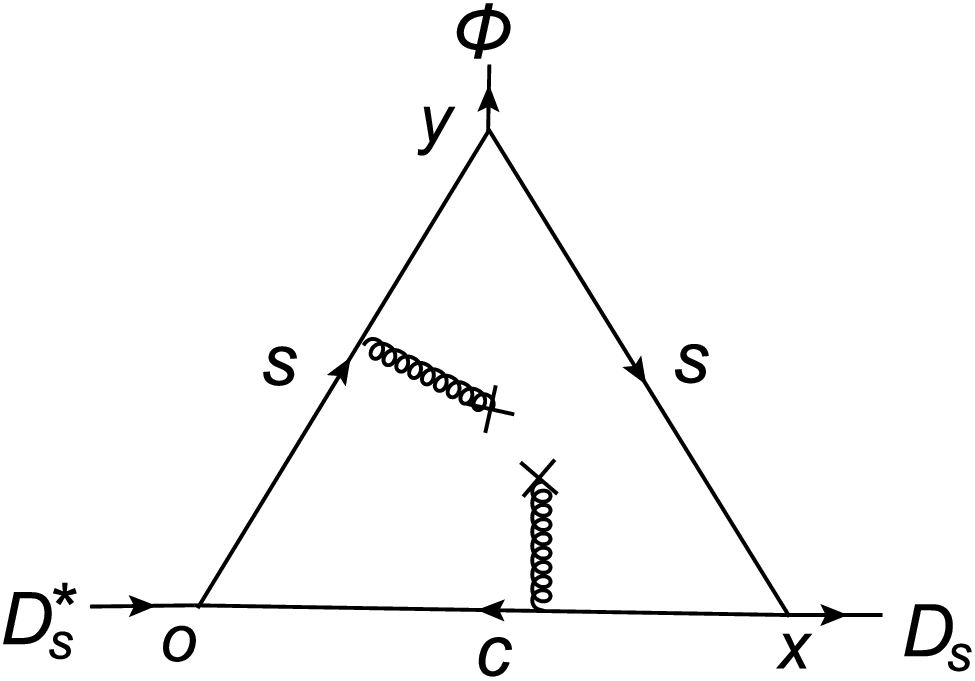}}
\subfigure[]{\includegraphics[height=2.0cm,width=2.6cm]{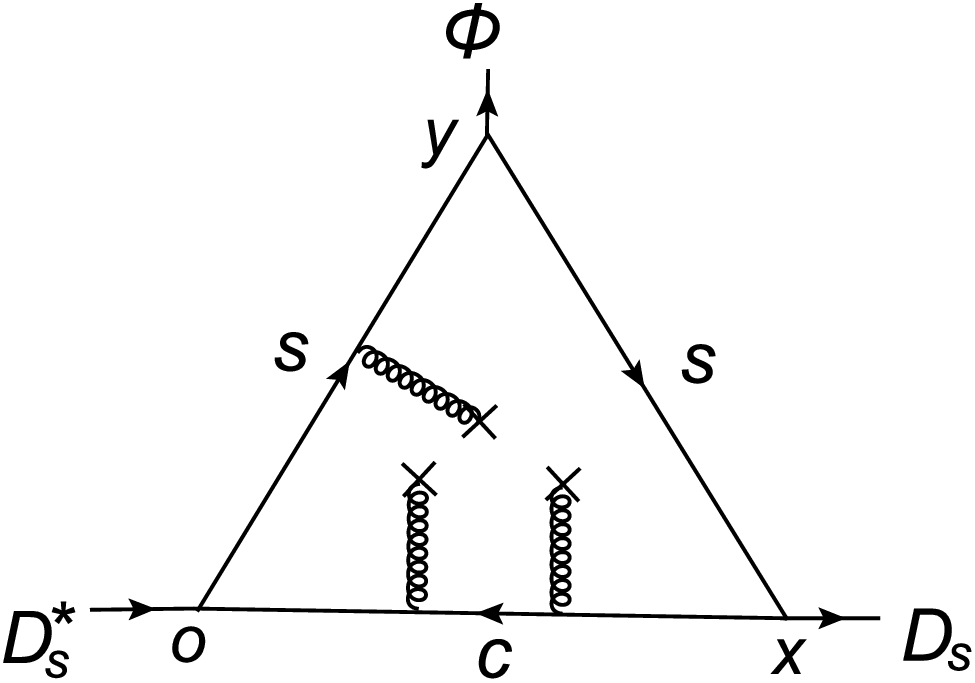}}
\subfigure[]{\includegraphics[height=2.0cm,width=2.6cm]{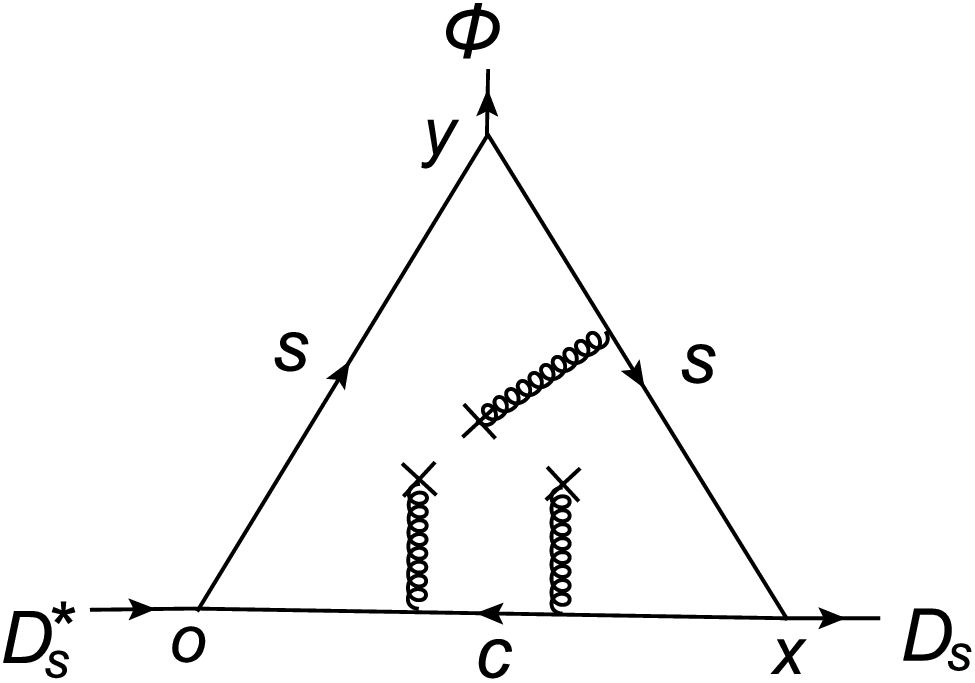}}

\subfigure[]{\includegraphics[height=2.0cm,width=2.6cm]{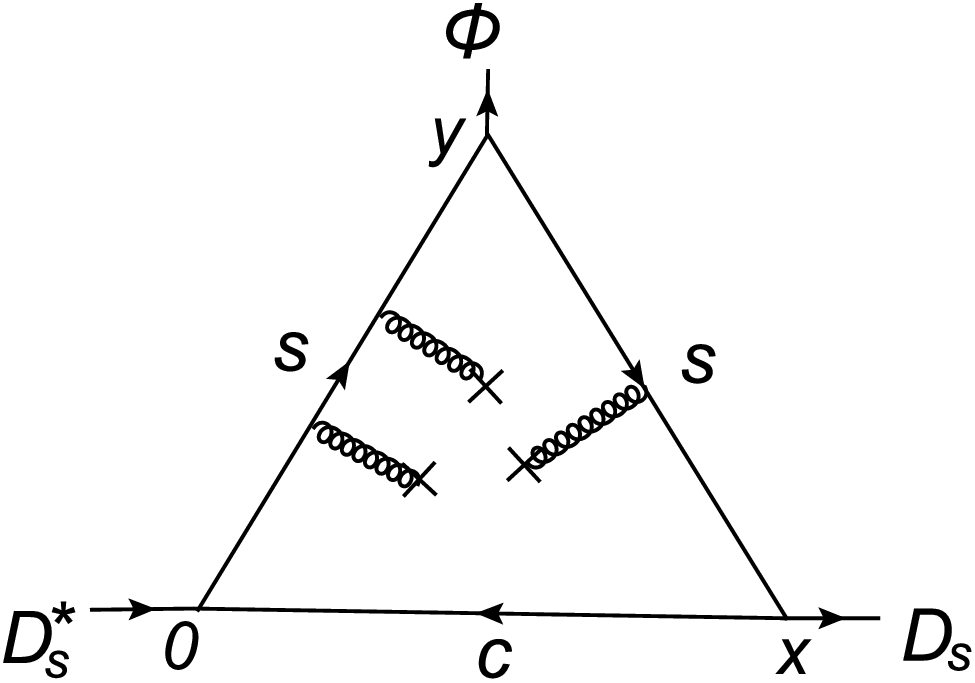}}
\subfigure[]{\includegraphics[height=2.0cm,width=2.6cm]{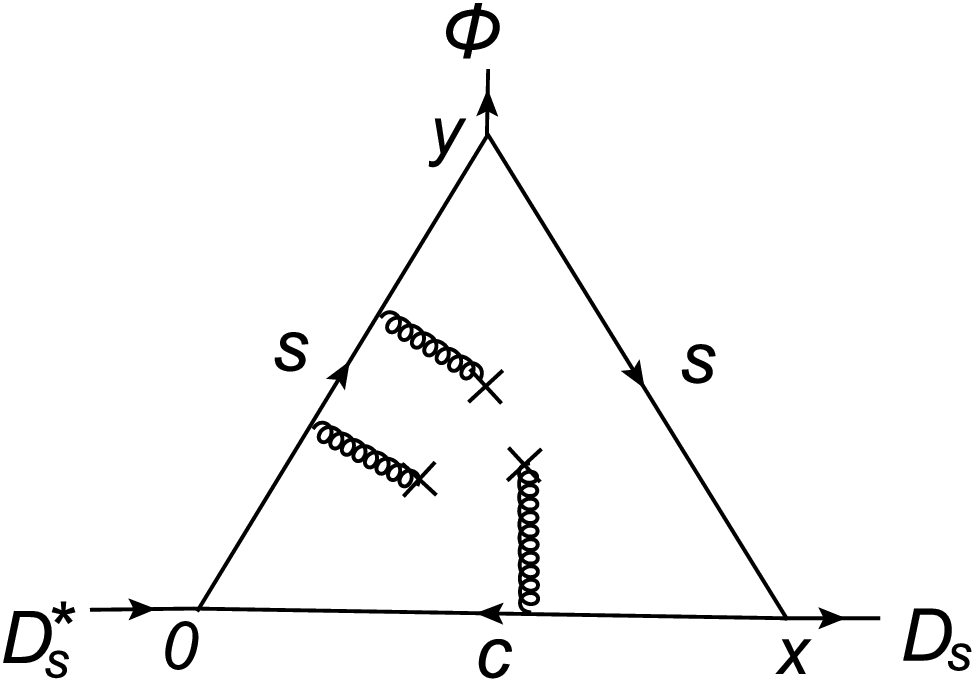}}
\subfigure[]{\includegraphics[height=2.0cm,width=2.6cm]{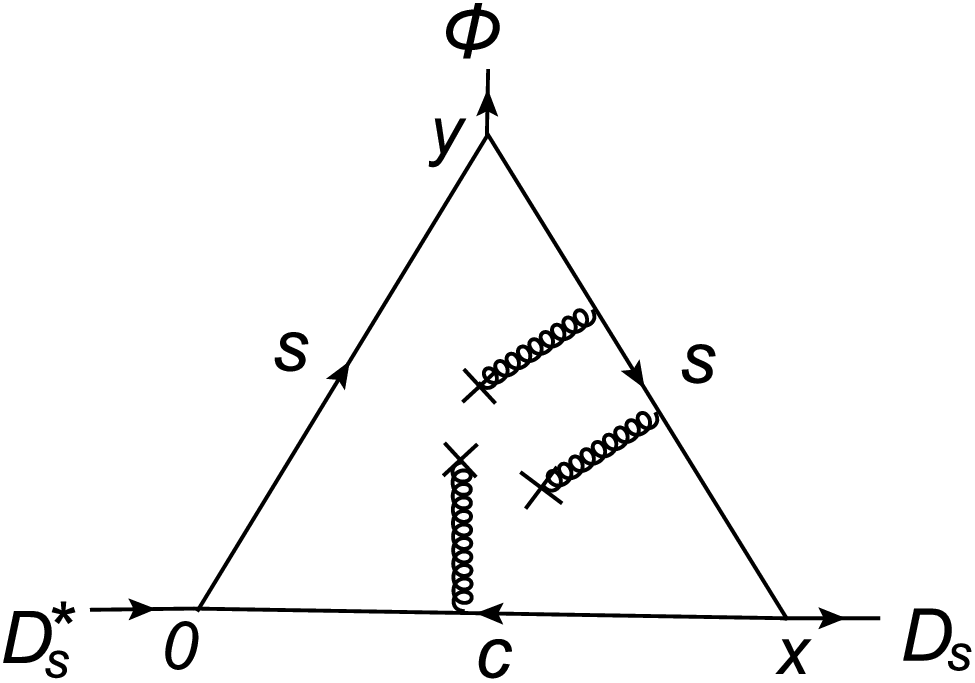}}
\subfigure[]{\includegraphics[height=2.0cm,width=2.6cm]{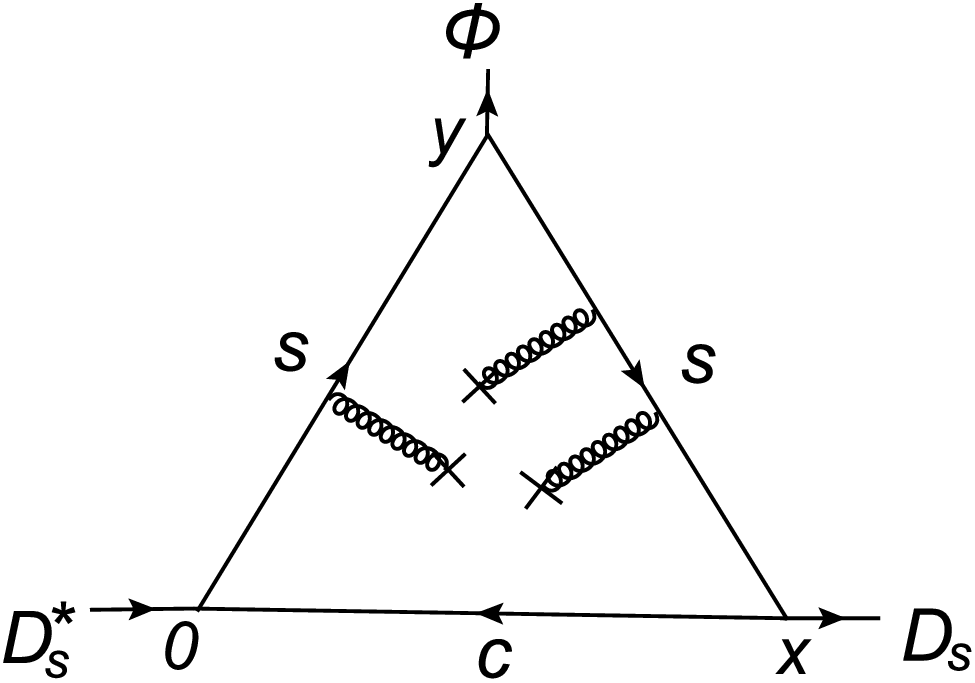}}

\subfigure[]{\includegraphics[height=2.0cm,width=2.6cm]{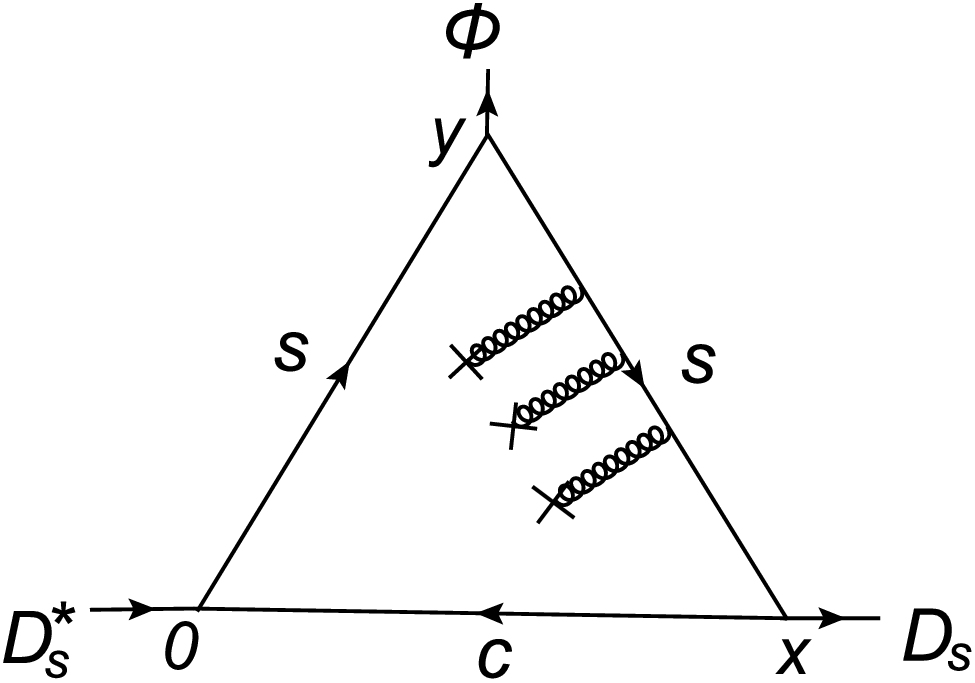}}
\subfigure[]{\includegraphics[height=2.0cm,width=2.6cm]{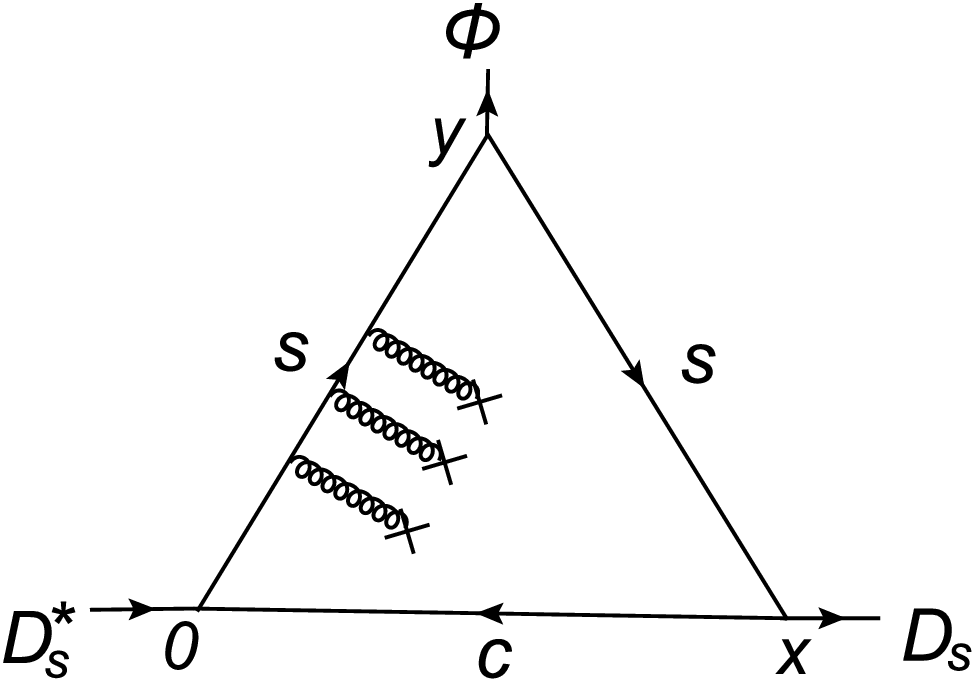}}
\subfigure[]{\includegraphics[height=2.0cm,width=2.6cm]{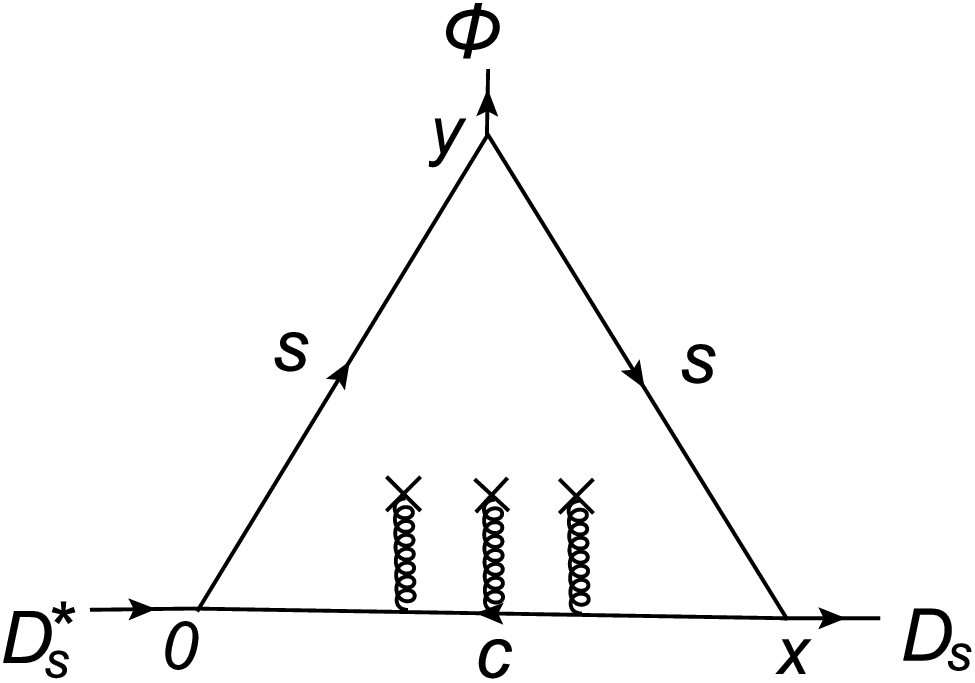}}
\subfigure[]{\includegraphics[height=2.0cm,width=2.6cm]{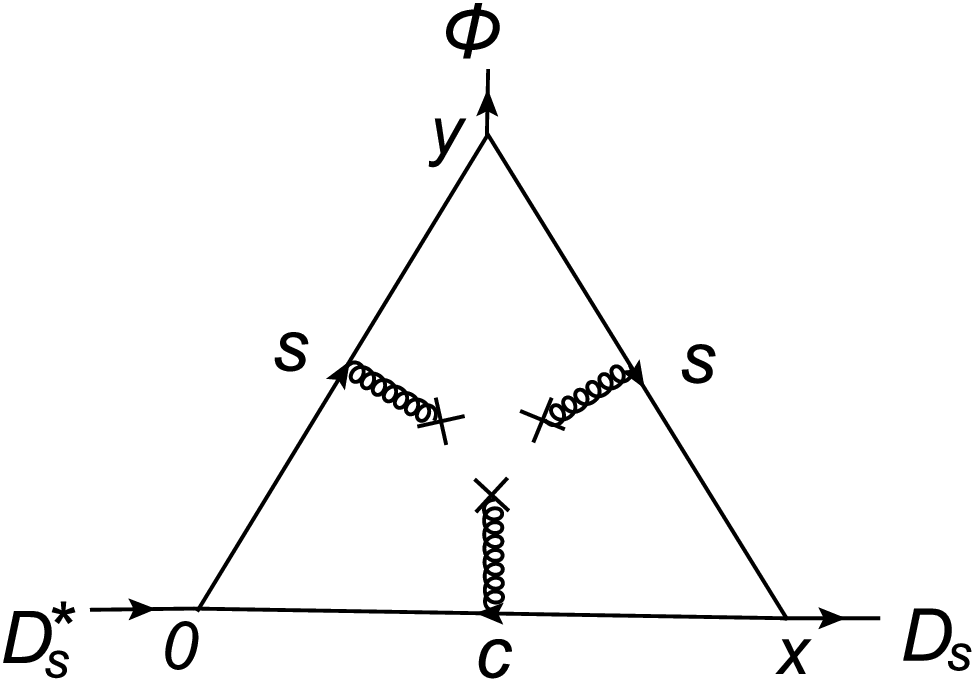}}
\end{center}
\caption{Contributions of the condensate parts $\langle
g^2G^2\rangle$ and $\langle f^3G^3\rangle$ for $\phi$ off-shell case
}
\end{figure}
\begin{figure}[htp]
\begin{center}
\subfigure[]{\includegraphics[height=2.0cm,width=2.6cm]{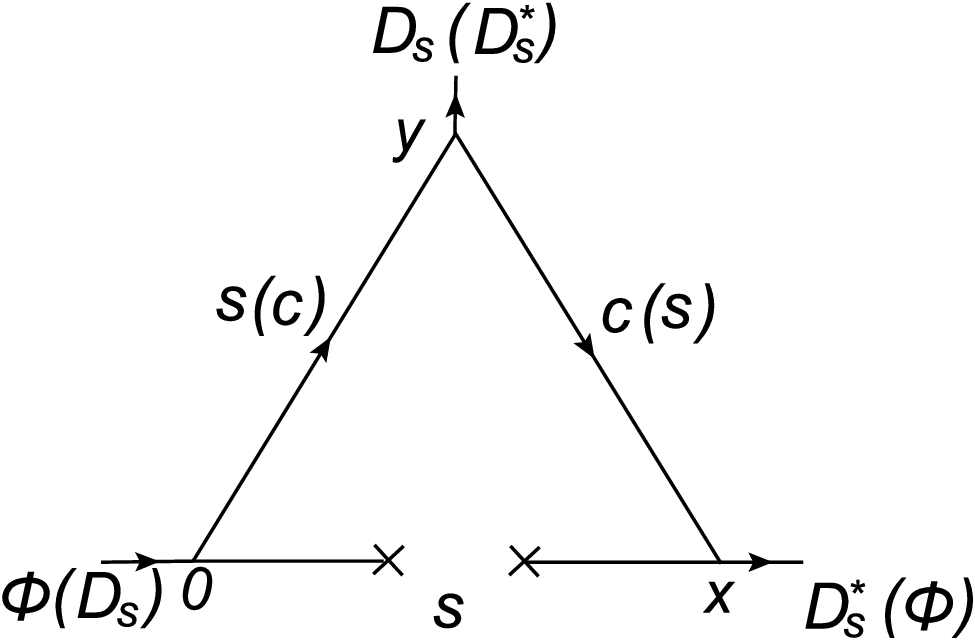}}
\subfigure[]{\includegraphics[height=2.0cm,width=2.6cm]{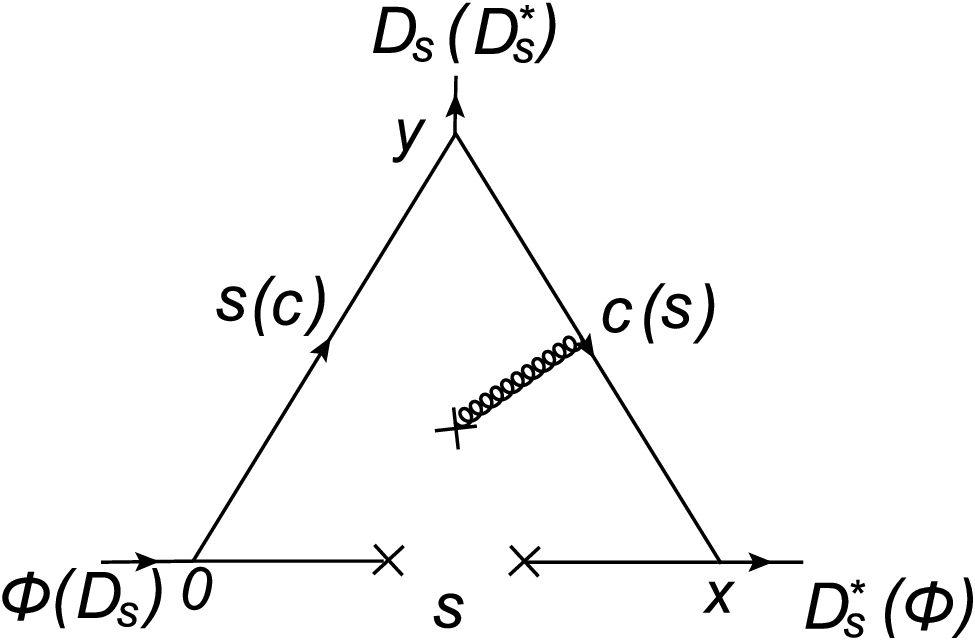}}
\subfigure[]{\includegraphics[height=2.0cm,width=2.6cm]{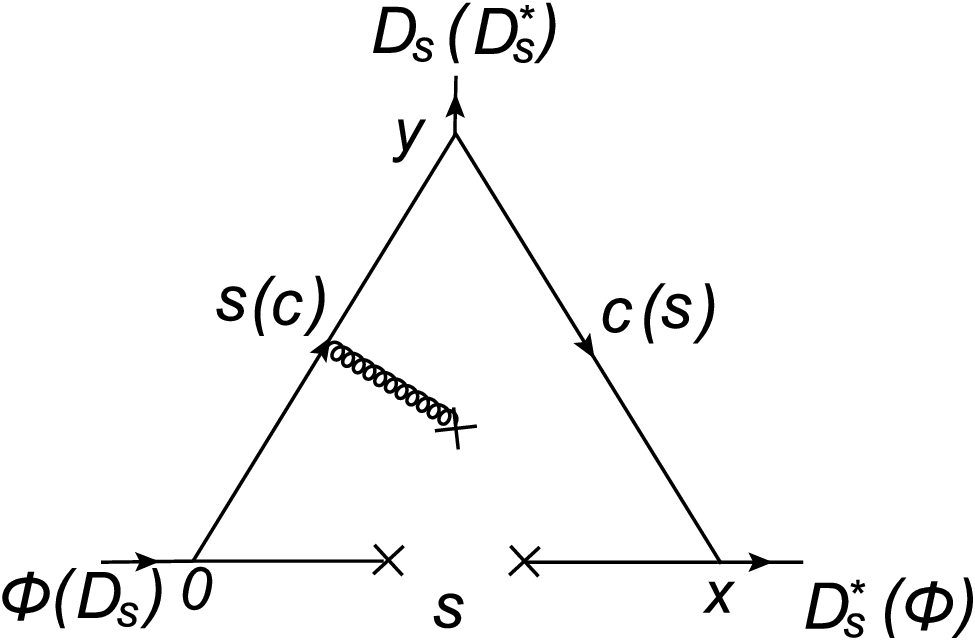}}
\subfigure[]{\includegraphics[height=2.0cm,width=2.6cm]{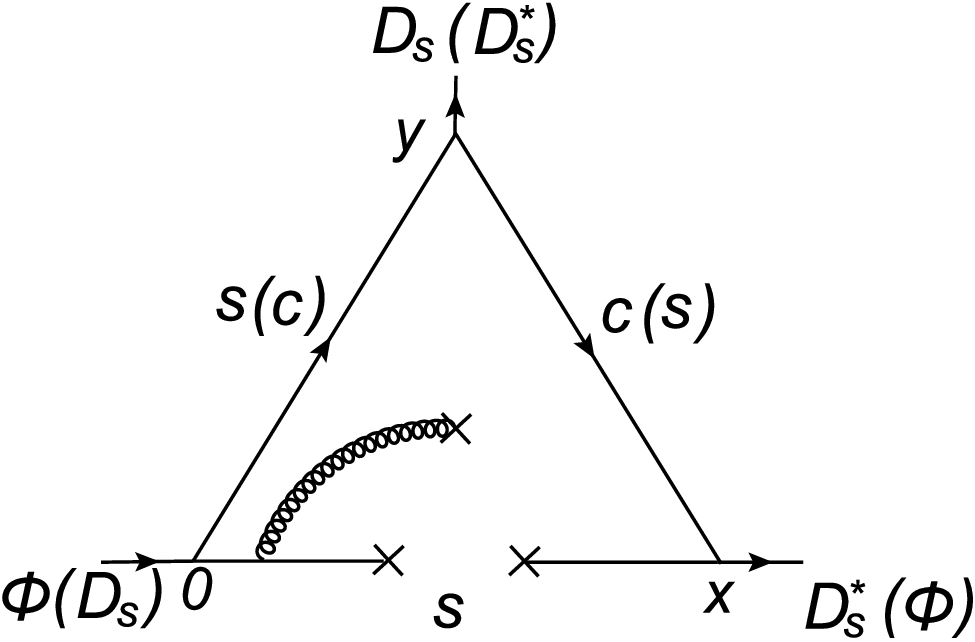}}
\subfigure[]{\includegraphics[height=2.0cm,width=2.6cm]{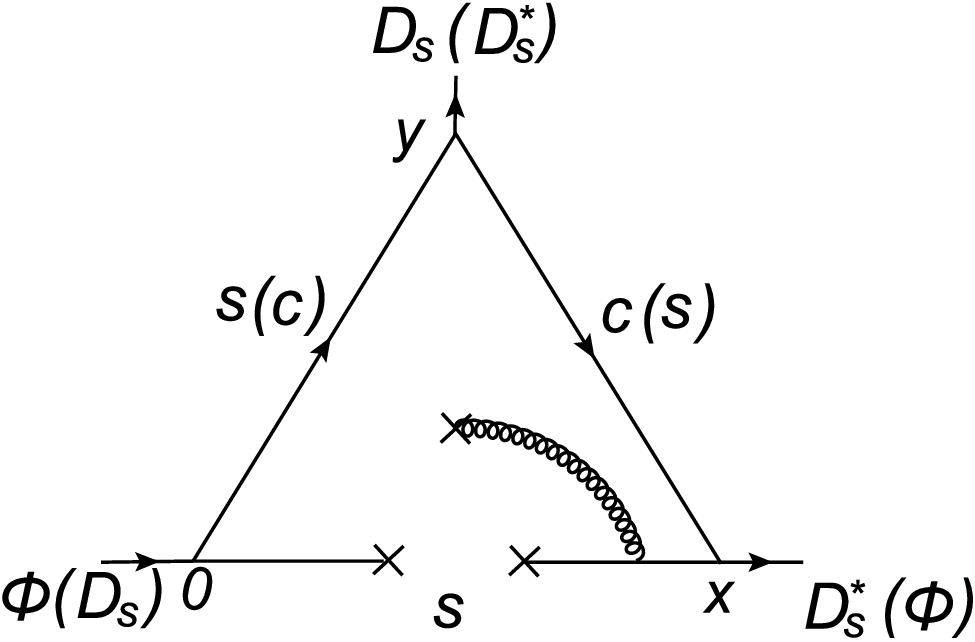}}

\subfigure[]{\includegraphics[height=2.0cm,width=2.6cm]{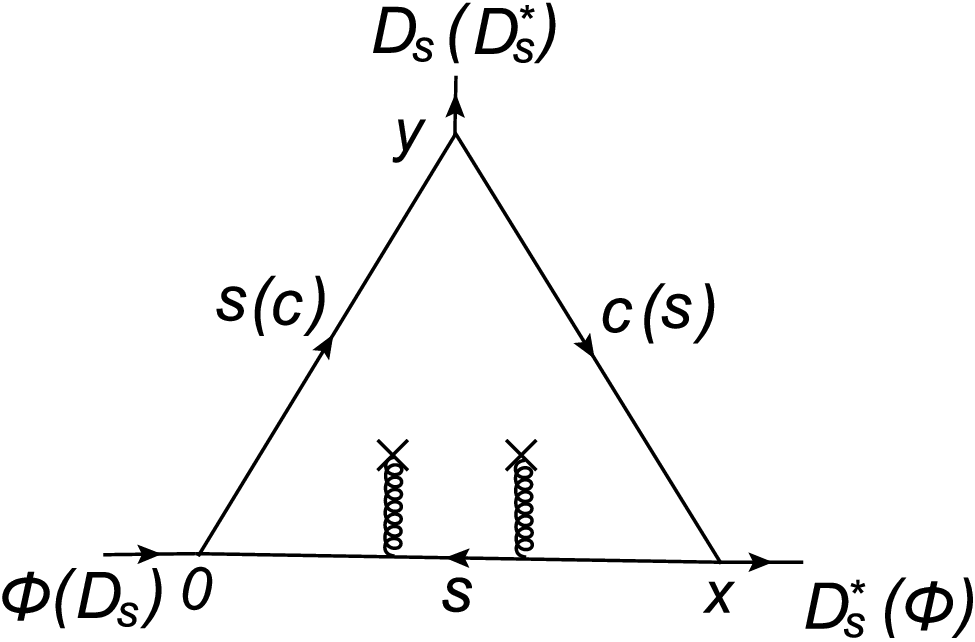}}
\subfigure[]{\includegraphics[height=2.0cm,width=2.6cm]{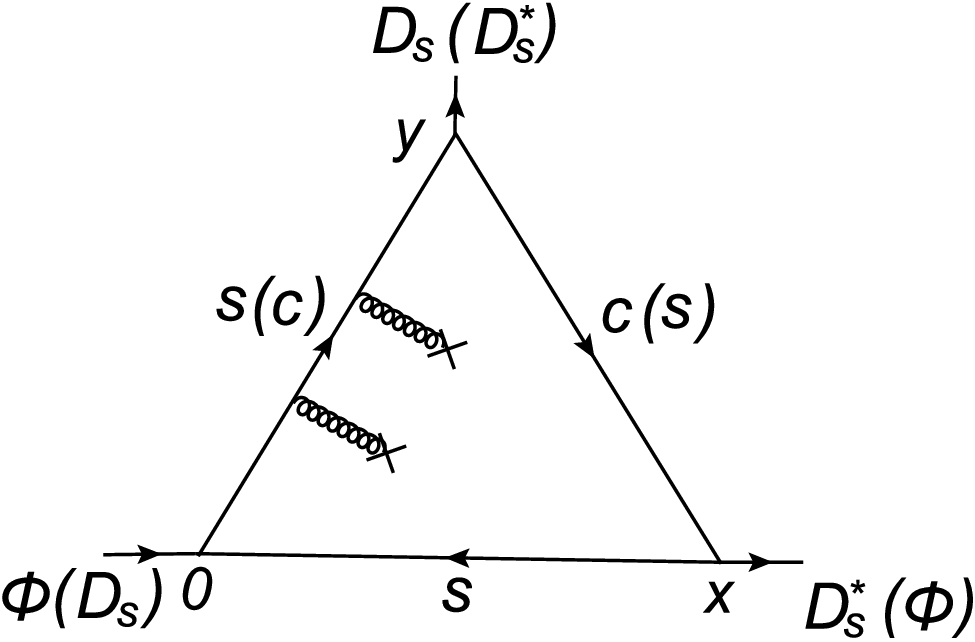}}
\subfigure[]{\includegraphics[height=2.0cm,width=2.6cm]{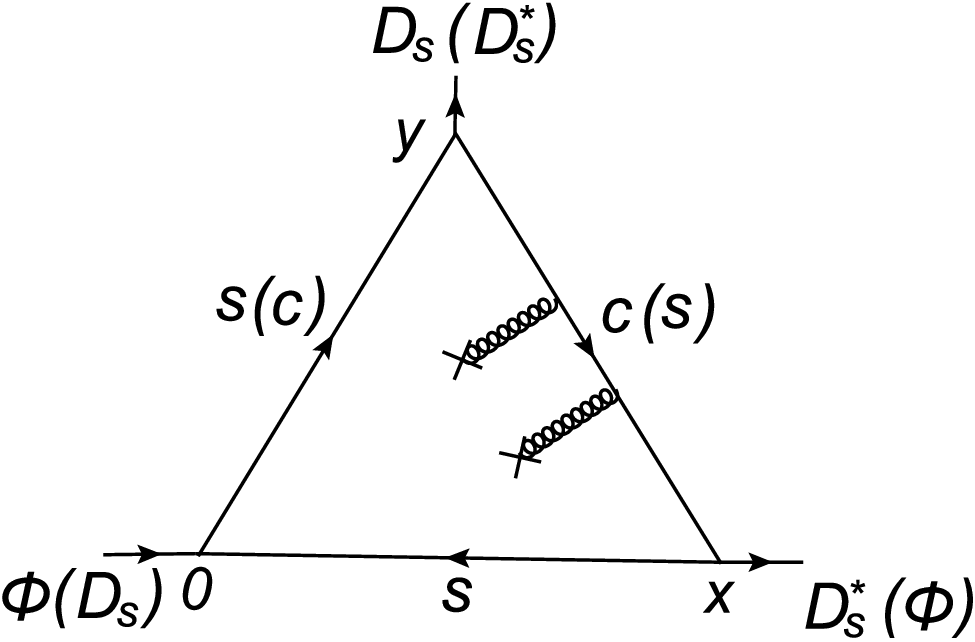}}
\subfigure[]{\includegraphics[height=2.0cm,width=2.6cm]{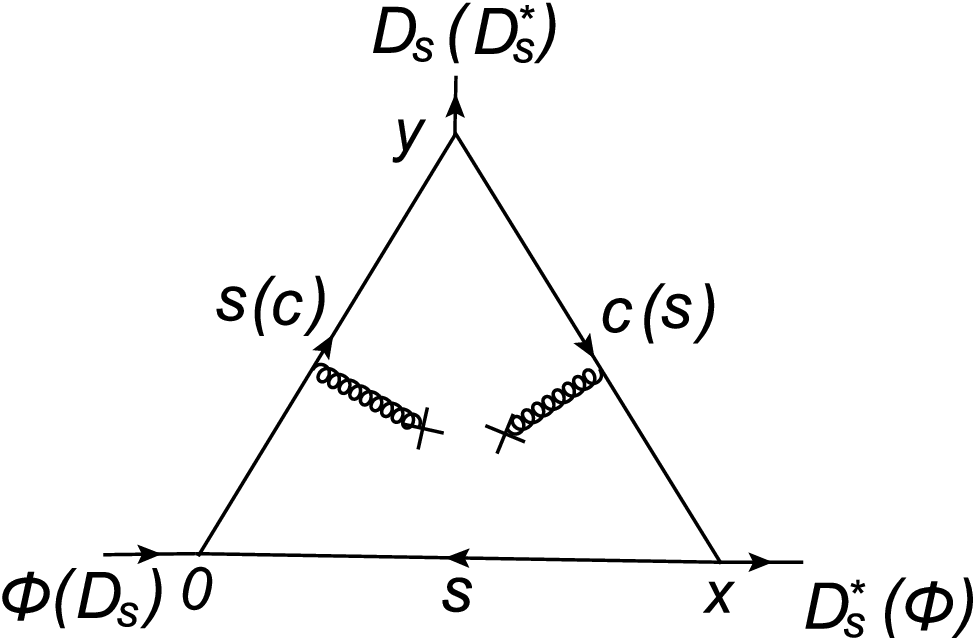}}
\subfigure[]{\includegraphics[height=2.0cm,width=2.6cm]{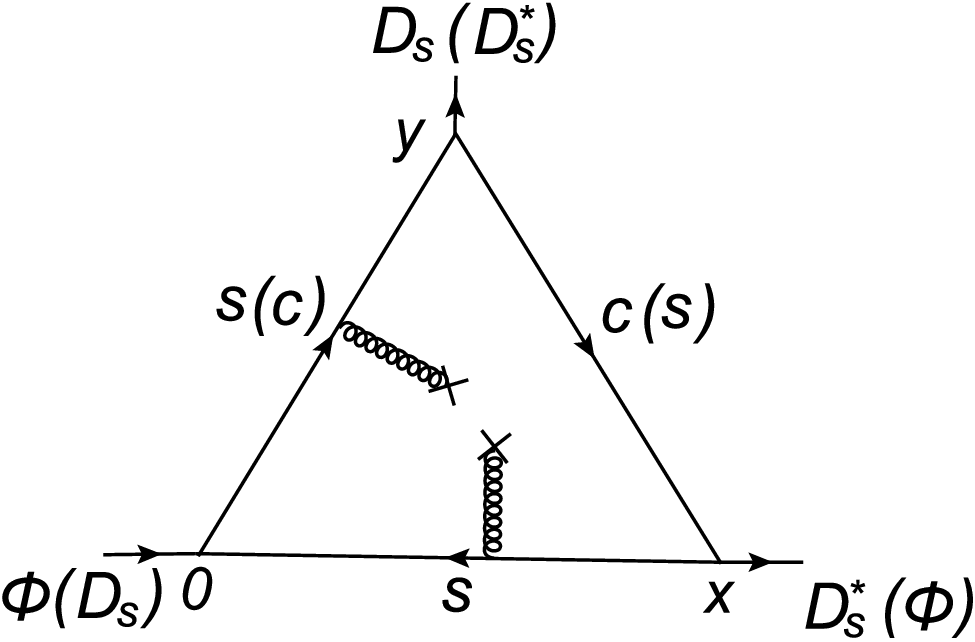}}

\subfigure[]{\includegraphics[height=2.0cm,width=2.6cm]{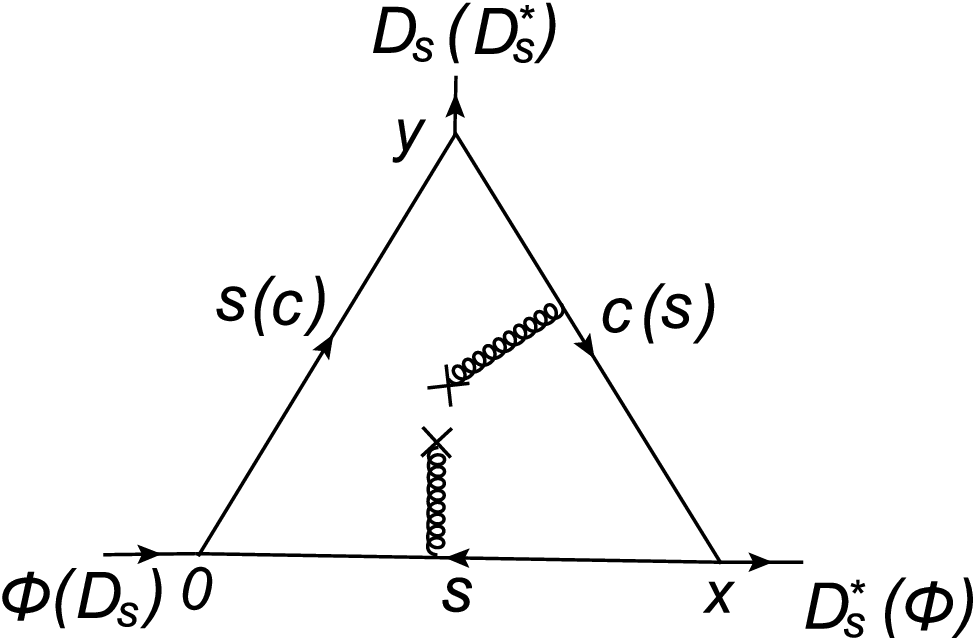}}
\subfigure[]{\includegraphics[height=2.0cm,width=2.6cm]{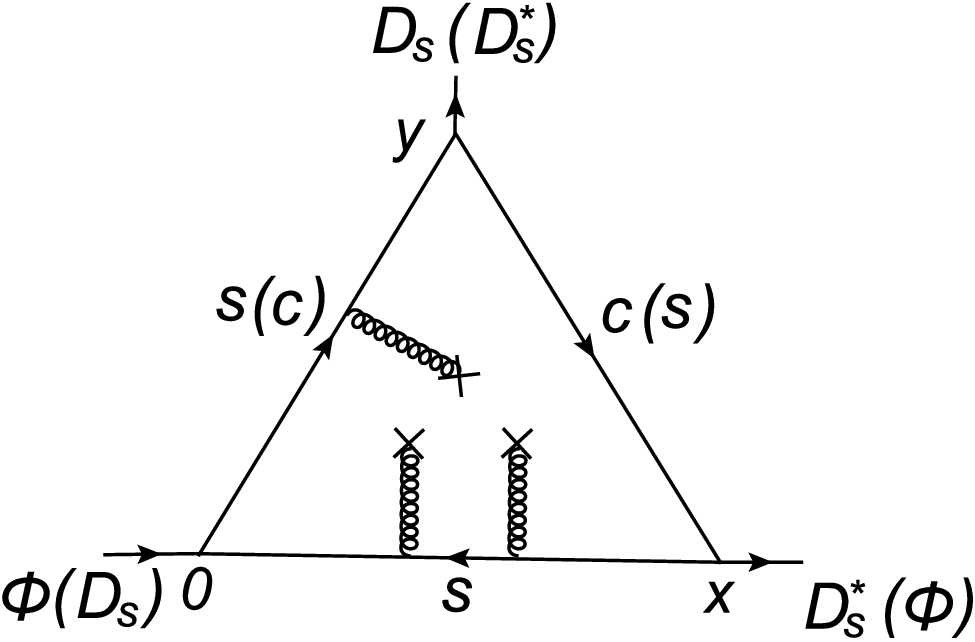}}
\subfigure[]{\includegraphics[height=2.0cm,width=2.6cm]{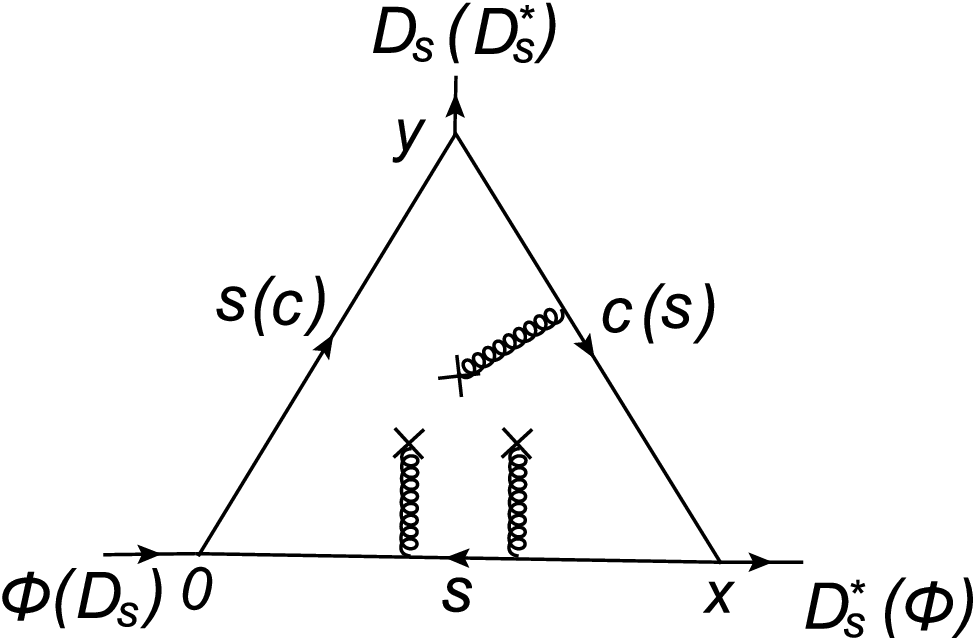}}
\subfigure[]{\includegraphics[height=2.0cm,width=2.6cm]{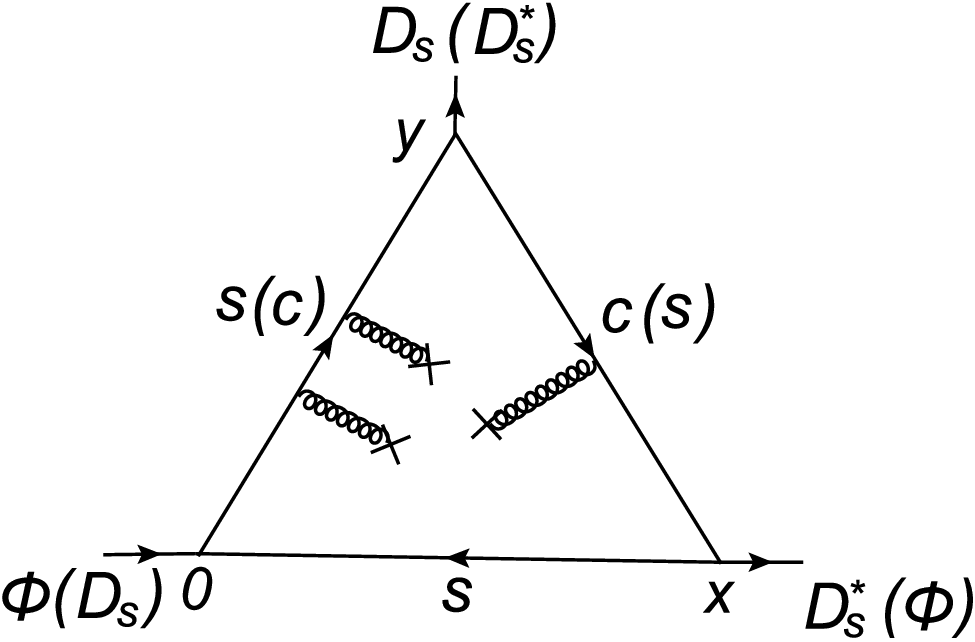}}
\subfigure[]{\includegraphics[height=2.0cm,width=2.6cm]{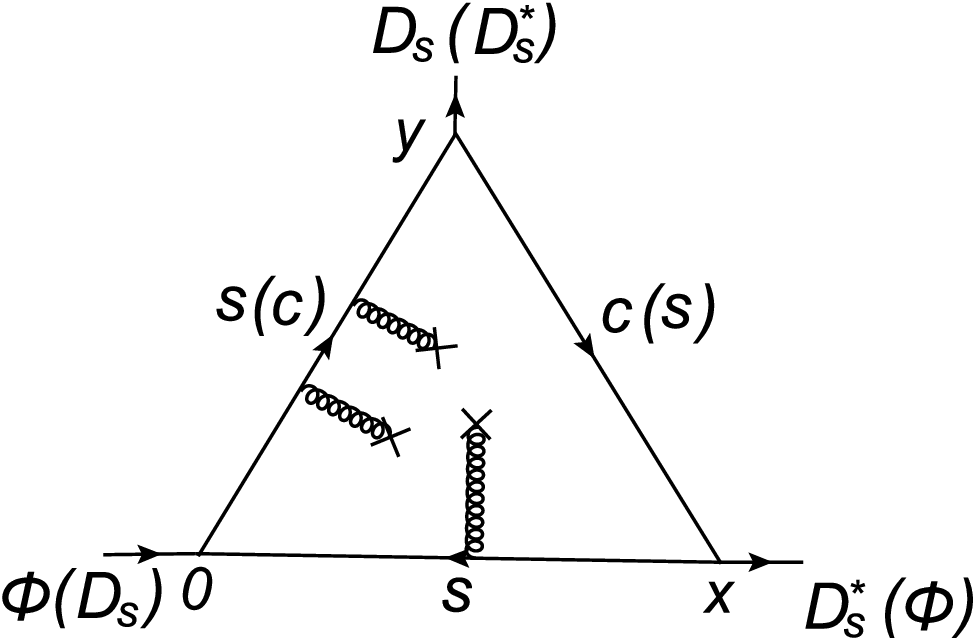}}

\subfigure[]{\includegraphics[height=2.0cm,width=2.6cm]{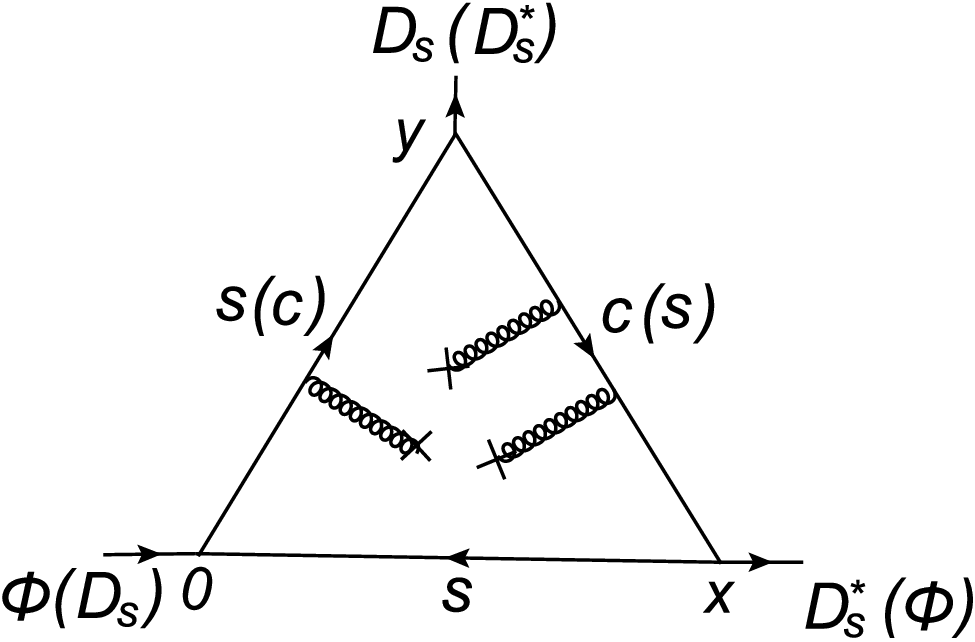}}
\subfigure[]{\includegraphics[height=2.0cm,width=2.6cm]{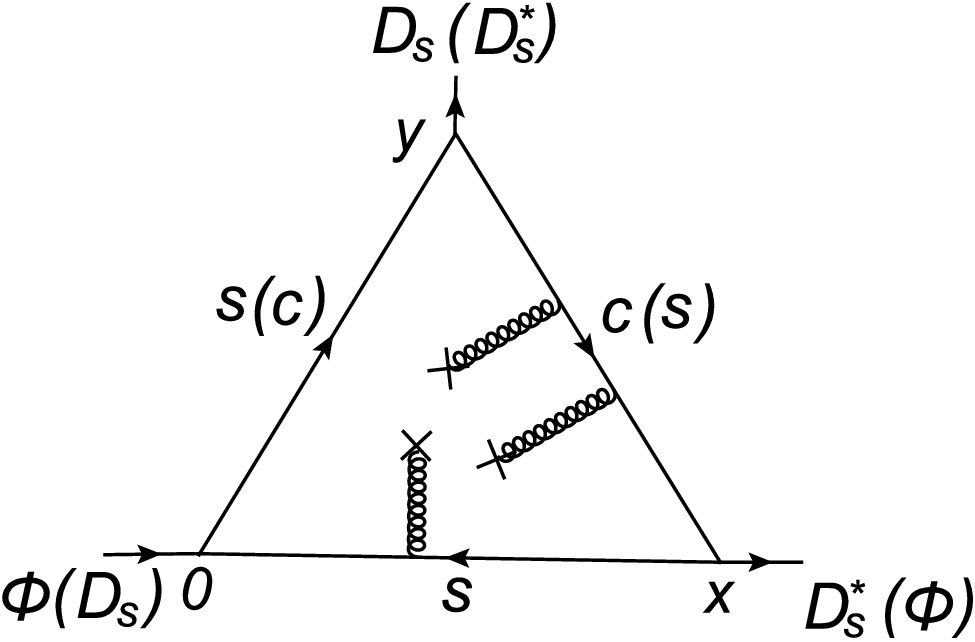}}
\subfigure[]{\includegraphics[height=2.0cm,width=2.6cm]{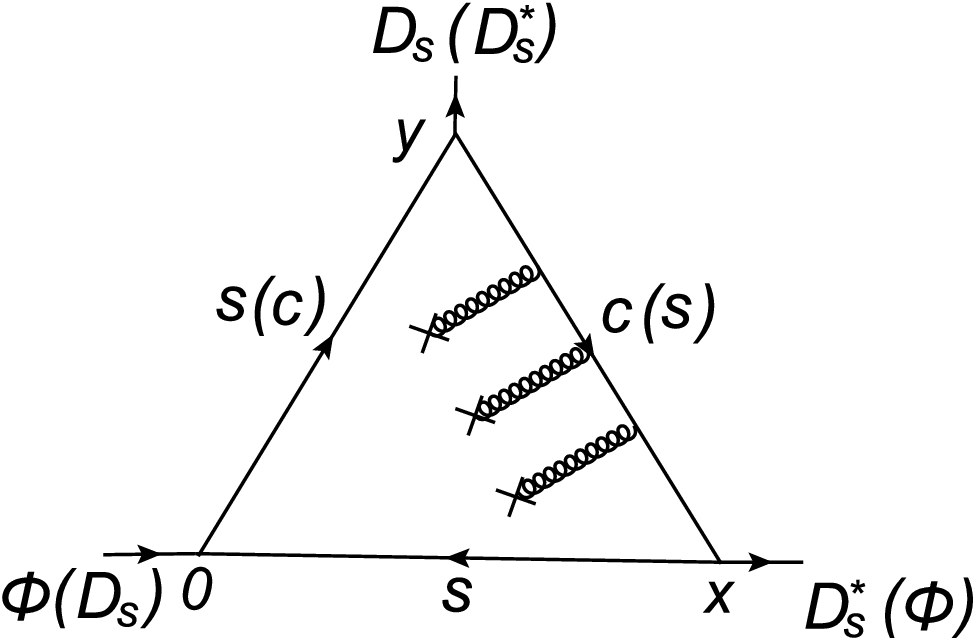}}
\subfigure[]{\includegraphics[height=2.0cm,width=2.6cm]{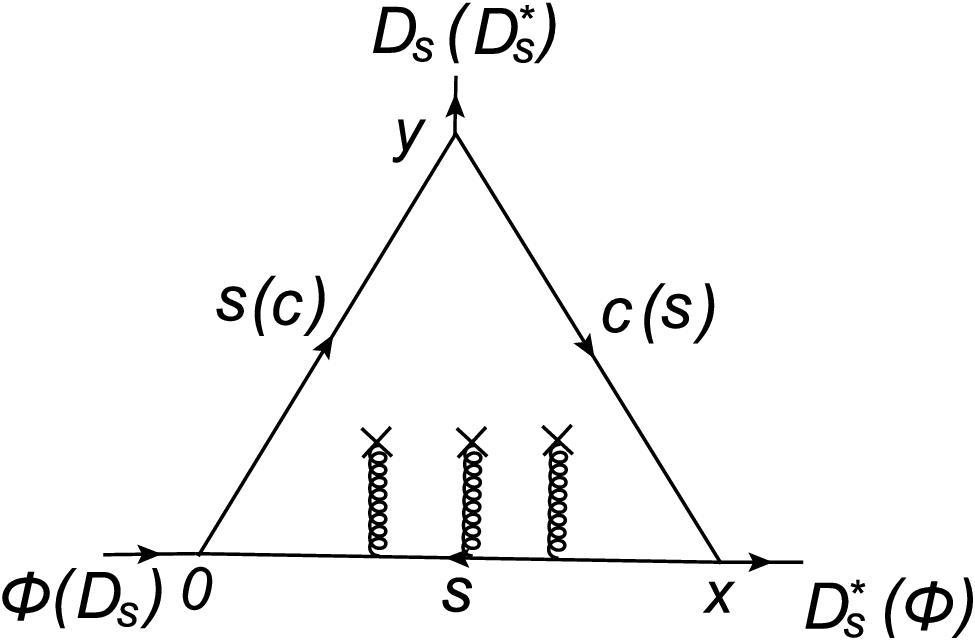}}
\subfigure[]{\includegraphics[height=2.0cm,width=2.6cm]{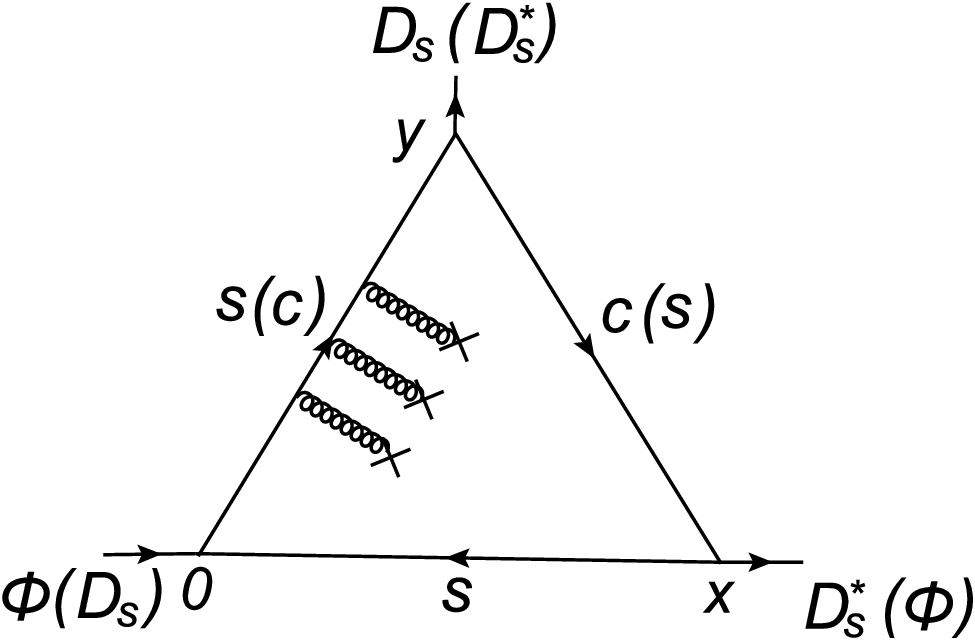}}

\subfigure[]{\includegraphics[height=2.0cm,width=2.6cm]{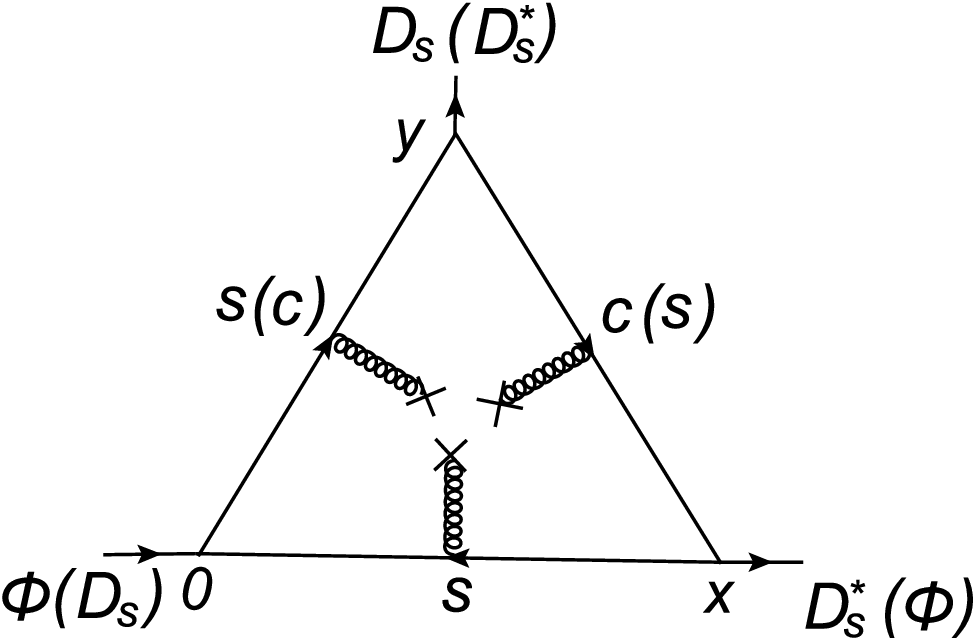}}
\subfigure[]{\includegraphics[height=2.0cm,width=2.6cm]{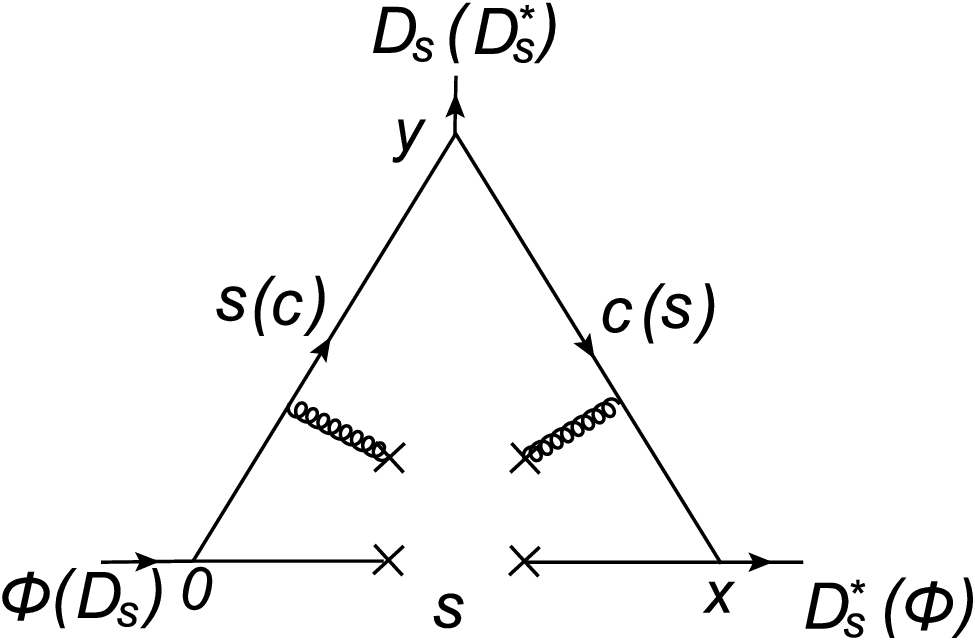}}
\subfigure[]{\includegraphics[height=2.0cm,width=2.6cm]{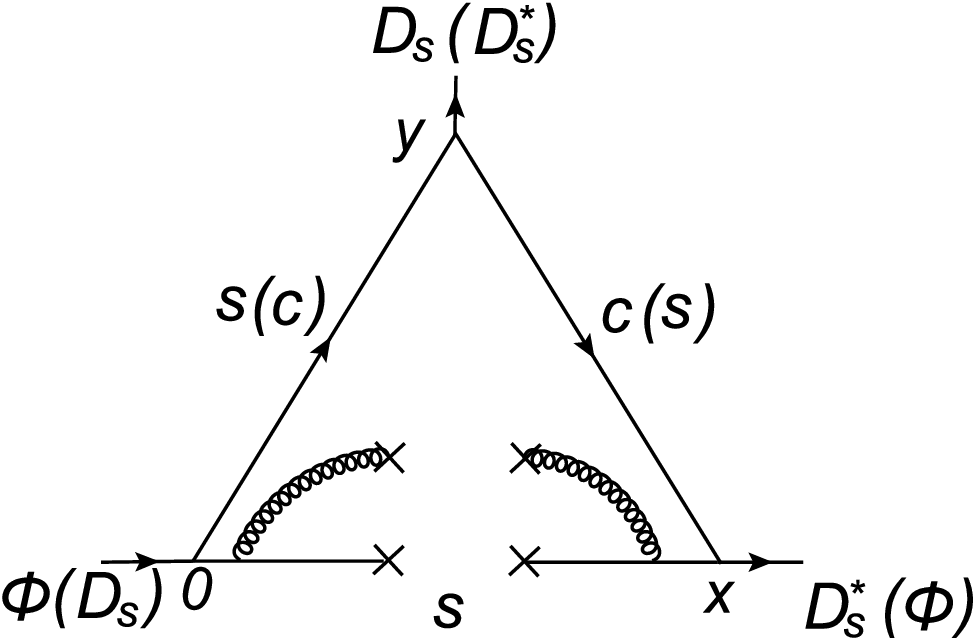}}
\subfigure[]{\includegraphics[height=2.0cm,width=2.6cm]{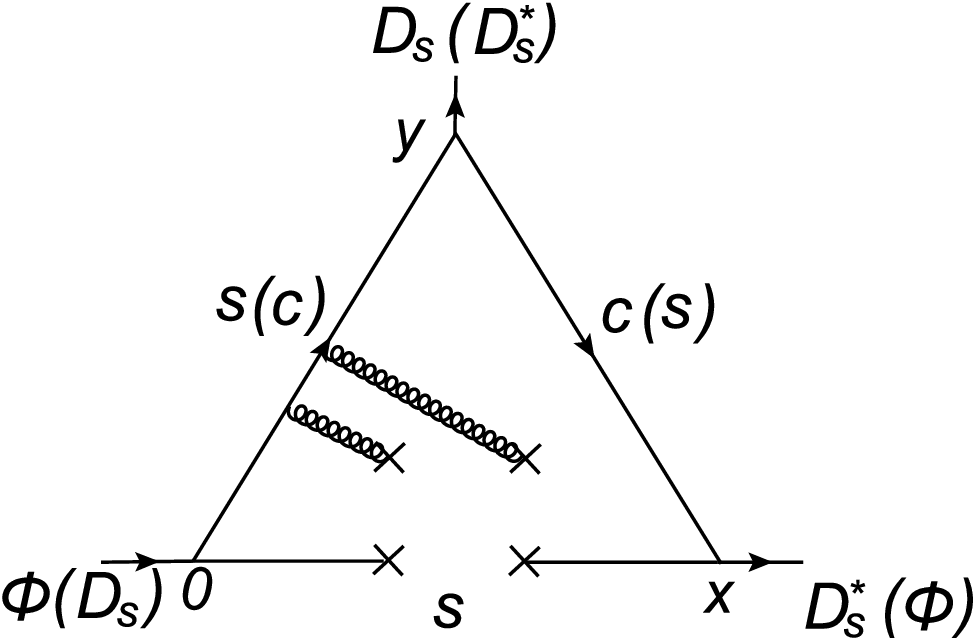}}
\subfigure[]{\includegraphics[height=2.0cm,width=2.6cm]{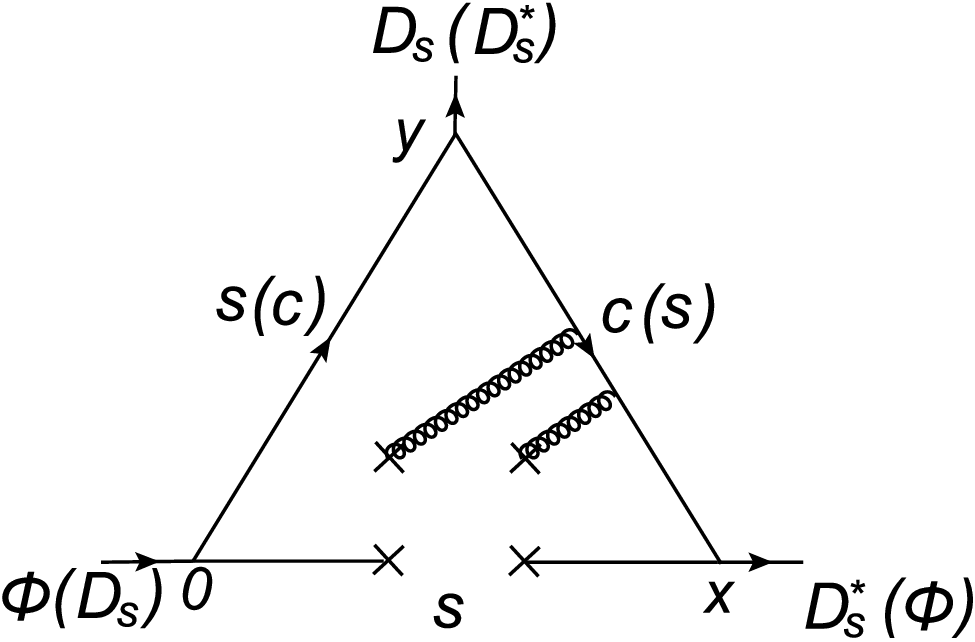}}

\subfigure[]{\includegraphics[height=2.0cm,width=2.6cm]{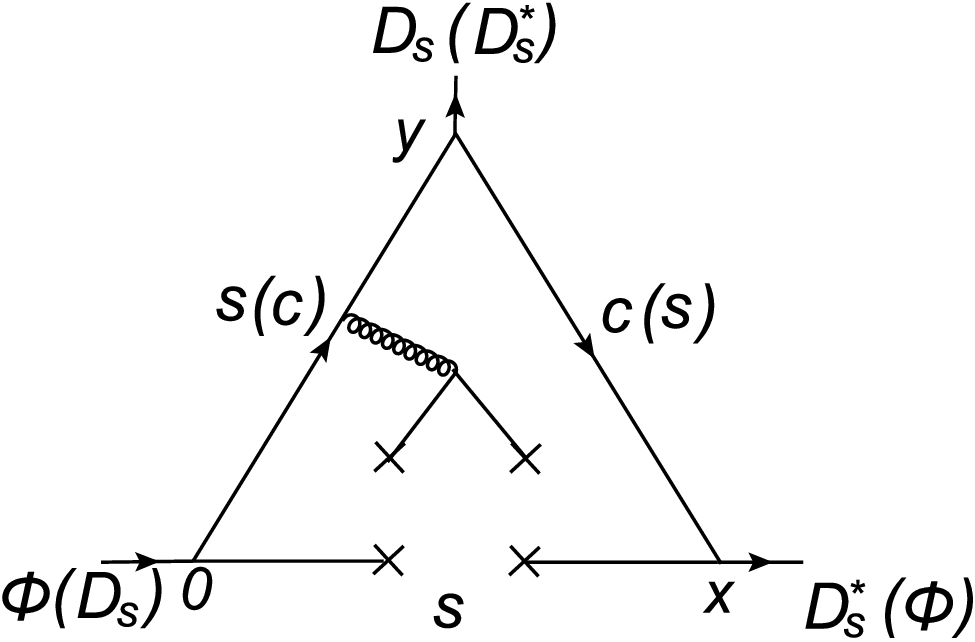}}
\subfigure[]{\includegraphics[height=2.0cm,width=2.6cm]{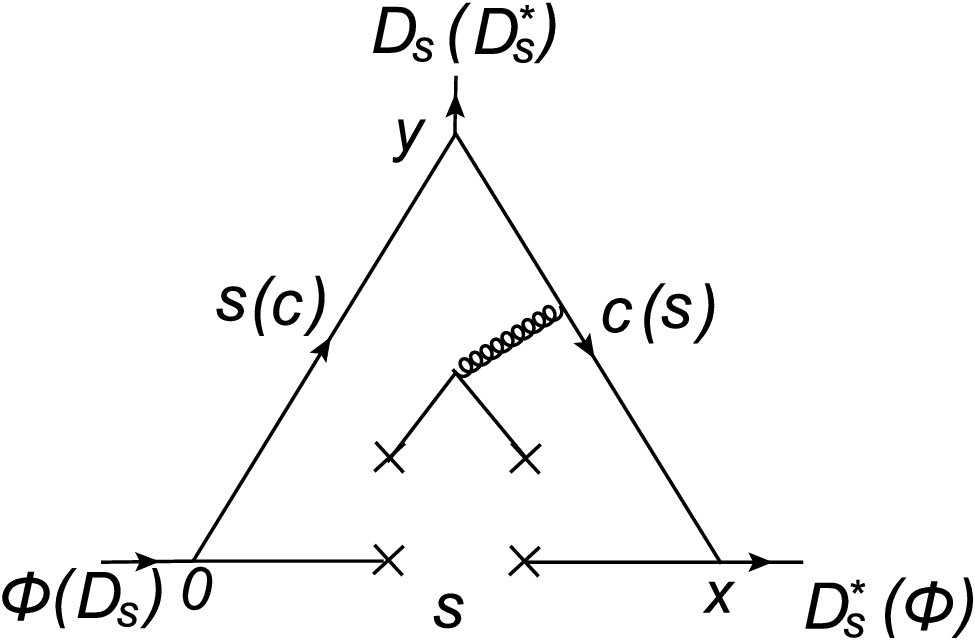}}
\subfigure[]{\includegraphics[height=2.0cm,width=2.6cm]{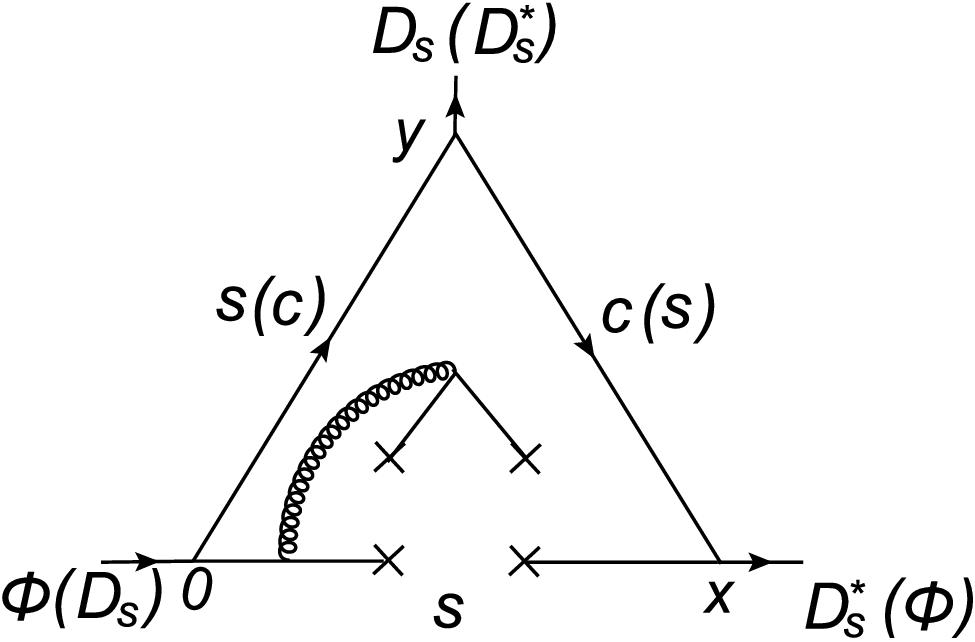}}
\subfigure[]{\includegraphics[height=2.0cm,width=2.6cm]{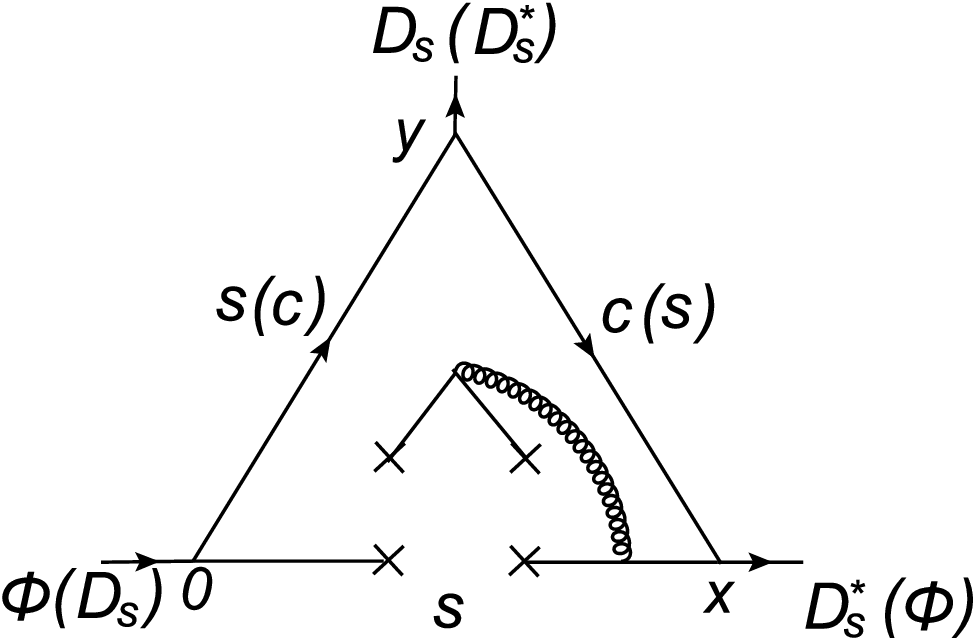}}
\end{center}
\caption{Contributions of the non-perturbative parts for
$D_{s}(D_{s}^{*})$ off-shell case}
\end{figure}

\begin{large}
\textbf{2.3 The Coupling Constant and the Meson decay}
\end{large}

We make the change of variables $p^{2}\rightarrow
-P^{2}$,$p'^{2}\rightarrow -P'^{2}$ and $q^{2}\rightarrow -Q^{2}$
and perform a double Borel transform\cite{Cola00} to the physical as
well as the OPE sides, which involves the transformation:
$P^{2}\rightarrow M_{1}^{2}$ and $P'^{2}\rightarrow M_{2}^{2}$,
where $M_{1}$ and $M_{2}$ are the Borel parameters. Then, we equate
the phenomenological and OPE sides, invoking the quark-hadron
duality from which the sum rule is obtained.

In order to eliminate the $h.r.$ terms from the phenomenological
side in Eqs.(7)$\sim(9)$, two continuum threshold parameters $s_{0}$
and $u_{0}$ in the OPE side are introduced. These parameters fulfill
the following relations: $m_{i}^{2}<s_{0}<m_{i}'^{2}$ and
$m_{o}^{2}<u_{0}<m_{o}'^{2}$, where $m_{i}$ and $m_{o}$ are the
masses of the incoming and out-coming mesons respectively and $m'$
is the mass of the first excited state of these mesons. After these
performaions, the form factors can be written as:

\begin{eqnarray}
G_{D_{s}^{\ast }D_{s}\phi}^{\phi }(Q^{2})&=&\frac{-\frac{1}{4\pi
^{2}M_{1}^{2}M_{2}^{2}}\int_{s_{1}}^{s_{0}}\int_{u_{1}}^{u_{0}}\rho
^{pert(\phi )}(s,u,Q^{2})e^{-\frac{s}{M_{1}^{2}}}e^{-\frac{u}{M_{2}^{2}}%
}dsdu+\mathscr{BB}\big[\Pi ^{non-pert(\phi
)}\big]}{\frac{C}{(Q^{2}+m_{\phi
}^{2})M_{1}^{2}M_{2}^{2}}e^{-m_{D_{s}^{\ast
}}^{2}/M_{1}^{2}}e^{-m_{D_{s}}^{2}/M_{2}^{2}}},
\end{eqnarray}
\begin{eqnarray}
G_{D_{s}^{\ast }D_{s}\phi}^{D_{s}}(Q^{2})&=&\frac{-\frac{1}{4\pi
^{2}M_{1}^{2}M_{2}^{2}}\int_{s_{1}}^{s_{0}}\int_{u_{1}}^{u_{0}}\rho
^{pert(D_{s})}(s,u,Q^{2})e^{-\frac{s}{M_{1}^{2}}}e^{-\frac{u}{M_{2}^{2}}%
}dsdu+\mathscr{BB}\big[\Pi ^{non-pert(D_{s})}\big]}{\frac{C}{%
(Q^{2}+m_{D_{s}}^{2})M_{1}^{2}M_{2}^{2}}e^{-m_{\phi
}^{2}/M_{1}^{2}}e^{-m_{D_{s}^{\ast }}^{2}/M_{2}^{2}}},
\end{eqnarray}
\begin{eqnarray}
G_{D_{s}^{\ast }D_{s}\phi}^{D_{s}^{\ast
}}(Q^{2})&=&\frac{-\frac{1}{4\pi
^{2}M_{1}^{2}M_{2}^{2}}\int_{s_{1}}^{s_{0}}\int_{u_{1}}^{u_{0}}\rho
^{pert(D_{s}^{\ast })}(s,u,Q^{2})e^{-\frac{s}{M_{1}^{2}}}e^{-\frac{u}{%
M_{2}^{2}}}dsdu+\mathscr{BB}\big[\Pi ^{non-pert(D_{s}^{\ast })}\big]}{\frac{C}{%
(Q^{2}+m_{D_{s}^{\ast }}^{2})M_{1}^{2}M_{2}^{2}}%
e^{-m_{D_{s}}^{2}/M_{1}^{2}}e^{-m_{\phi }^{2}/M_{2}^{2}}},
\end{eqnarray}
where $\mathscr{BB}[\;\;]$ stands for the double Borel transform.
Now, we can calculate the form factors in the space-like region
according to these above equations. However, in order to obtain the
coupling constants, it is necessary to extrapolate these results
into physical regions($Q^{2}<0$), which is realized by fit the form
factors into suitable analytical functions. It is indicated that we
should get the same values for the coupling constants
$G^{\phi}_{D_{s}^{*}D_{s}\phi}$, $G^{D_{s}}_{D_{s}^{*}D_{s}\phi}$
and $G^{D_{s}^{*}}_{D_{s}^{*}D_{s}\phi}$\cite{Brra01}, when we take
$Q^{2}=-m_{\phi}^{2}$, $Q^{2}=-m_{D_{s}}^{2}$ and
$Q^{2}=-m_{D_{s}^{*}}^{2}$ separately. This above procedure is used
to minimize the uncertainties in the calculation of the coupling
constant, which will be quite clear in the following section.

With the assumption of the vector meson dominance($\phi(1020)$), the
radiative decays $D_{s}^{*}\rightarrow D_{s}\gamma$ can be described
by the following electromagnetic lagrangian $\pounds'$,
\begin{eqnarray}
\pounds'=-eQ_{s}\overline{s}\gamma_{\mu}sA^{\mu}
\end{eqnarray}
where the $A_{\mu}$, $Q_{s}$ are the electromagnetic field and the
charge number. From the lagrangian $\pounds'$, we can obtain the
decay amplitude\cite{Wang10},
\begin{eqnarray}
\notag\ && \left\langle D_{s}(p)\gamma (q,\varepsilon )|D_{s}^{\ast
}(p^{\prime },\xi )\right\rangle \\
\notag\
&=&\left\langle \gamma (q,\varepsilon )|\phi (q,\eta )\right\rangle \frac{i}{%
q^{2}-m_{\phi }^{2}}\left\langle D_{s}(p)\phi (q,\eta )|D_{s}^{\ast
}(p^{\prime },\xi )\right\rangle \\
\notag\ &=&\left\langle D_{s}(p)\phi (q,\eta )|D_{s}^{\ast
}(p^{\prime },\xi )\right\rangle \frac{i}{q^{2}-m_{\phi
}^{2}}f_{\phi }m_{\phi }eQ_{s}(-i)\varepsilon _{\mu }^{\ast }\eta
^{\mu }\\
&=&G_{D_{s}^{*}D_{s}\gamma}\epsilon ^{\alpha \beta \lambda \tau
}p_{\alpha }^{\prime }q_{\lambda }\xi _{\beta }\eta _{\tau }^{\ast
}\frac{i}{q^{2}-m_{\phi }^{2}}f_{\phi }m_{\phi
}eQ_{s}(-i)\varepsilon _{\mu }^{\ast }\eta ^{\mu }
\end{eqnarray}
The parameters $G_{D_{s}^{*}D_{s}\gamma}$ and $f_{\phi}$ are the
coupling constant and the weak decay constant, respectively.
$p_{\alpha}'$ and $q_{\lambda}$ are the four momenta of the $D_{s}$
and $\gamma$. $\eta^{\mu}$, $\varepsilon^{*}_{\mu}$ and
$\xi_{\beta}$ are the polarization vectors of the $\phi$, $\gamma$
and $D^{*}_{s}$, respectively. The strong coupling constant
$G_{D_{s}^{*}D_{s}\gamma}$ can be related to the effective coupling
constant in the heavy quark effective Lagrangian by Eq.(10) in this
paper.

\begin{large}
\textbf{3 Results and Discussions}
\end{large}

Present section is devoted to the numerical analysis of the sum
rules for the coupling constants. The decay constants and hadronic
parameters used in this work are taken as
$f_{\phi}=0.229\pm0.003$\cite{Oliv},
$f_{D_{s}}=0.257\pm0.006$\cite{Oliv},
$f_{D_{s}^{*}}=0.301\pm0.013$\cite{Oliv}, $m_{\phi}=1.019\pm0.020
GeV$\cite{Oliv}, $m_{D_{s}}=1.968\pm0.00032GeV$\cite{Oliv},
$m_{D_{s}^{*}}=2.112\pm0.0005GeV$\cite{Oliv}. The vacuum condensates
are taken to be the standard values
$<\overline{s}s>=-(0.8\pm0.1)\times(0.24\pm0.01GeV)^3$\cite{Cola00,Colan06},
$<\overline{s}g_{s}\sigma
Gs>=m_{0}^{2}<\overline{s}s>$\cite{Cola00,Colan06},
$m_{0}^{2}=(0.8\pm0.1)GeV^2$,
$<g_{s}^{2}GG>=(0.022\pm0.004)GeV^4$\cite{Nari10},
$<f^3G^3>=(8.8\pm5.5)GeV^{2}<g_{s}^{2}GG>$\cite{Nari10}. And we also
take the masses of quark $m_{c}=(1.275\pm0.025) GeV$,
$m_{s}=0.095\pm0.005GeV$ from the Particle Data Group\cite{Oliv}.
The continuum parameters, $s_{0}$ and $u_{0}$ in Eqs.(29)$\sim$(31),
are defined as $s_{0}=(m_{i}+\bigtriangleup_{i})^{2}$ and
$u_{0}=(m_{o}+\bigtriangleup_{o})^{2}$, where the quantities
$\bigtriangleup_{i}$ and $\bigtriangleup_{o}$ are determined
imposing the most stable Borel window. In order to include the pole
and to exclude the $h.r.$ contributions for the cases of $\phi$,
$D_{s}$ and $D_{s}^{*}$ mesons off-shell, the values for
$\triangle_{\phi}$, $\triangle_{D_{s}}$ and $\triangle_{D_{s}^{*}}$
can not be far from the experimental value of the distance between
the pole and the first excited state\cite{Cola00}. In addtion, the
results of the form factors in Eqs.(29)$\sim$(31) should also not
depend on the Borel parameters $M_{1}^{2}$ and $M_{2}^{2}$.
Therefore, we have to work in a region where the approximations made
are supposedly acceptable and where the results depend only
moderately on the Borel variables\cite{Cola00}. Using the Borel
region $5.0\leq M_{1}^{2}\leq7.0GeV^2$ and $5.0\leq
M_{2}^{2}\leq7.0GeV^2$ ($Q^2=3.0GeV^2$ for $\phi$ off-shell),
$6.0\leq M_{1}^{2}\leq8.0GeV^2$ and $6.0\leq M_{2}^{2}\leq8.0GeV^2$
($Q^2=1.0GeV^2$ for $D_{s}$ and $D_{s}^{*}$ off-shell) we found a
good stability with $\bigtriangleup_{\phi}=
\bigtriangleup_{D_{s}}=\bigtriangleup_{D_{s}^{*}}=0.5GeV$(Fig.4).

From the figure, we can see that the values are rather stalbe with
variations of the Borel parameters,  it is reliable to extract the
form factors. Besides of the pertubative term, we can also see that
$\langle s\overline{s}\rangle$ give a considerable contribution for
$D_{s}$ and $D_{s}^{*}$ off-shell cases(Fig.4 (c)$\sim$(f)). And the
contributions of the other condensate terms are small($<1\%$). To
the case of $\phi$ off-shell, condensate parts $\langle
g^2G^2\rangle$ and $\langle f^3G^3\rangle$ make up $ 1\%\sim2\% $ of
the total contributions. It should be noticed that although these
condensates terms, all except for the perturbative term and $\langle
s\overline{s}\rangle$, give small contributions to the form factors,
they have a significant influence on the following analytical
functions(Eqs.$(34)\sim(36)$), which are obtained by numerical
fitting. Thus, these condensates contributions should not be
neglected in the calculation.

\begin{figure}[htp]
\begin{center}
\subfigure[]{\includegraphics[height=5.0cm,width=6.6cm]{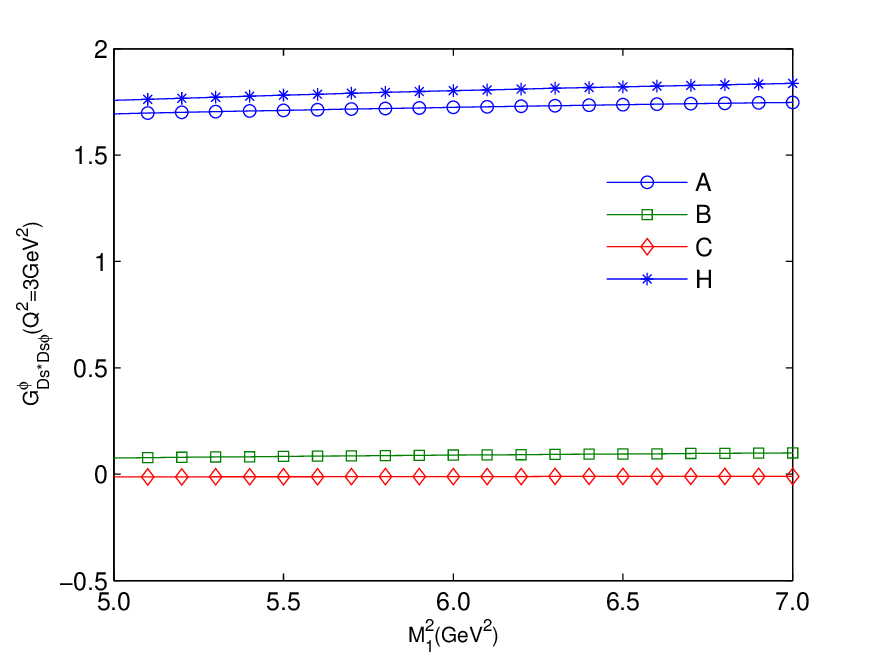}}
\subfigure[]{\includegraphics[height=5.0cm,width=6.6cm]{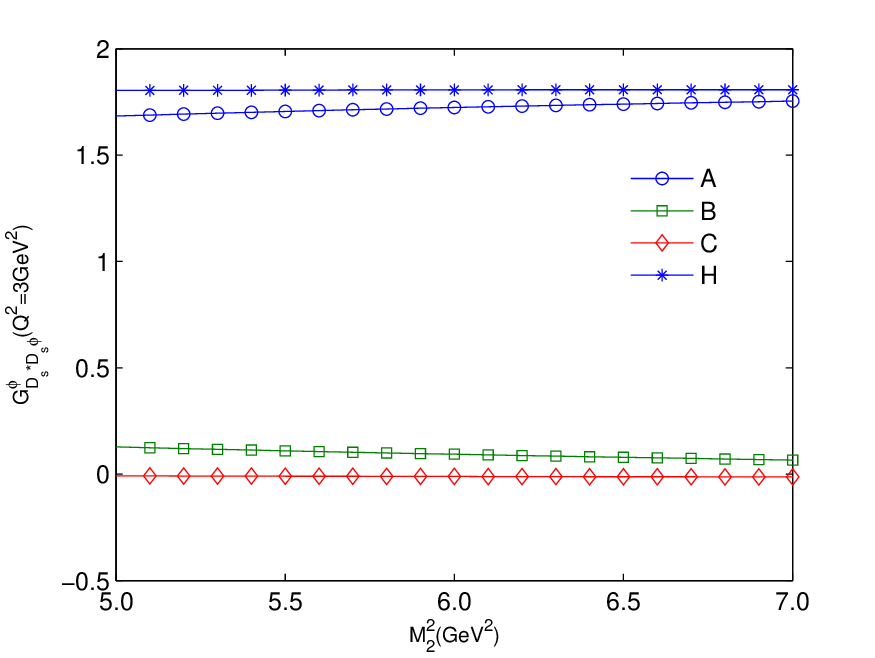}}
\subfigure[]{\includegraphics[height=5.0cm,width=6.6cm]{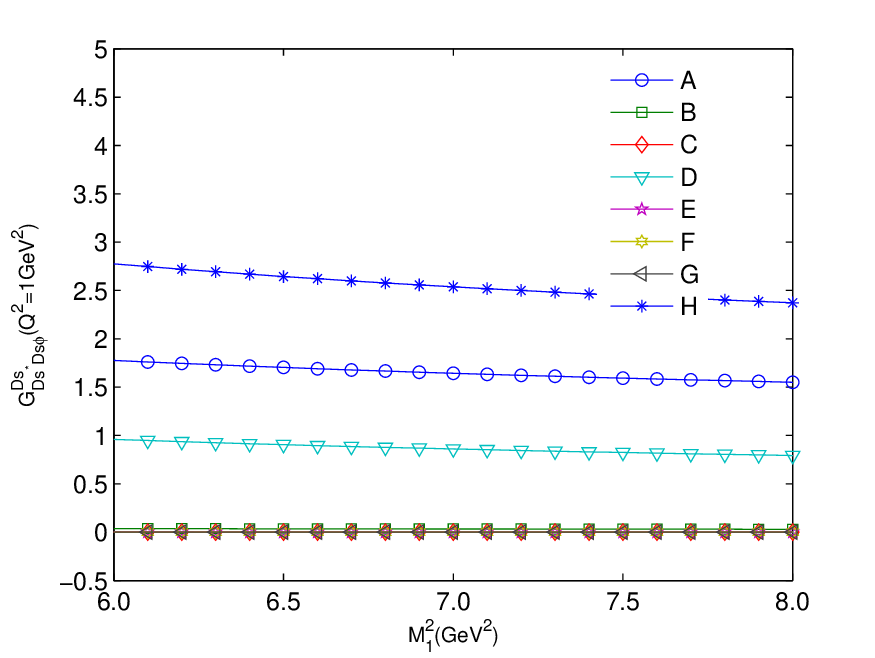}}
\subfigure[]{\includegraphics[height=5.0cm,width=6.6cm]{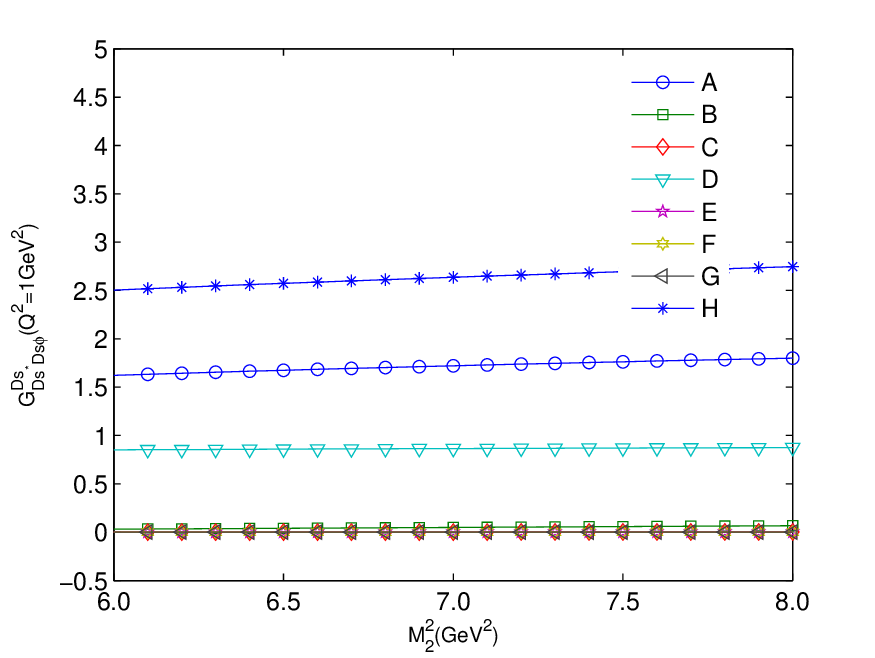}}
\subfigure[]{\includegraphics[height=5.0cm,width=6.6cm]{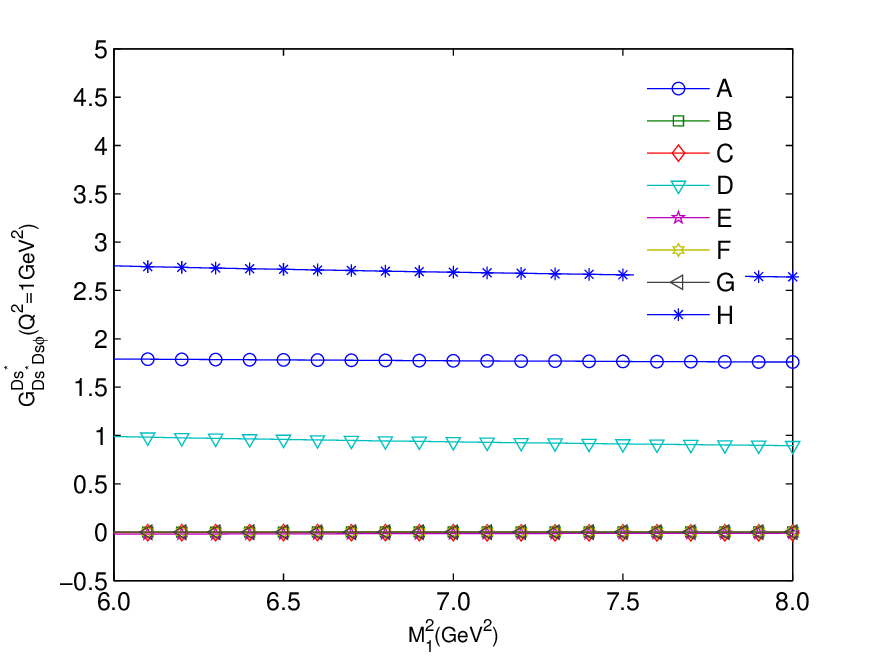}}
\subfigure[]{\includegraphics[height=5.0cm,width=6.6cm]{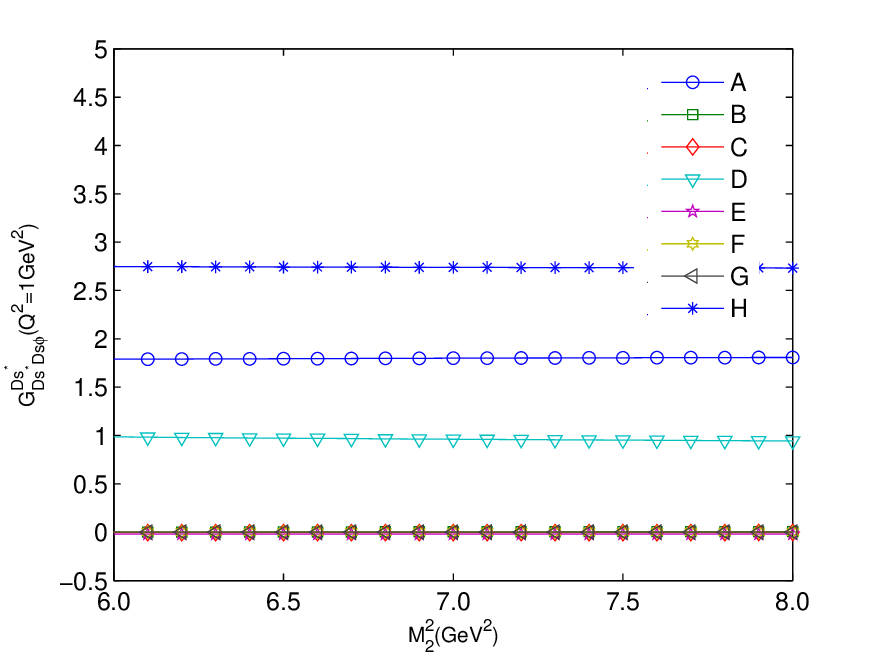}}
\end{center}
\caption{The contributions of different condensate terms in the OPE
with variations of the Borel parameters $M_{1}^{2}$ and $M_{2}^{2}$
for $\phi$((a),(b)), $D_{s}$((c),(d)) and $D_{s}^{*}$((e),(f))
off-shell, where A-H denote the perturbative term,$\langle
g^2G^2\rangle$, $\langle f^3G^3\rangle$, $\langle
s\overline{s}\rangle$, $\left\langle \overline{s}g\sigma
.Gs\right\rangle$,  $\langle s\overline{s} \rangle^2$,  $\langle
s\overline{s}\rangle \langle GG\rangle$ and Total contributions.}
\end{figure}

The form factors $G^{\phi}_{D_{s}^{*}D_{s}\phi}$,
$G^{D_{s}}_{D_{s}^{*}D_{s}\phi}$ and
$G^{D_{s}^{*}}_{D_{s}^{*}D_{s}\phi}$ are shown explicitly in Fig.5
and are fitted into the folowing analytical functions:

\begin{figure}[htp]
\begin{center}
\subfigure[]{\includegraphics[height=4.0cm,width=4.6cm]{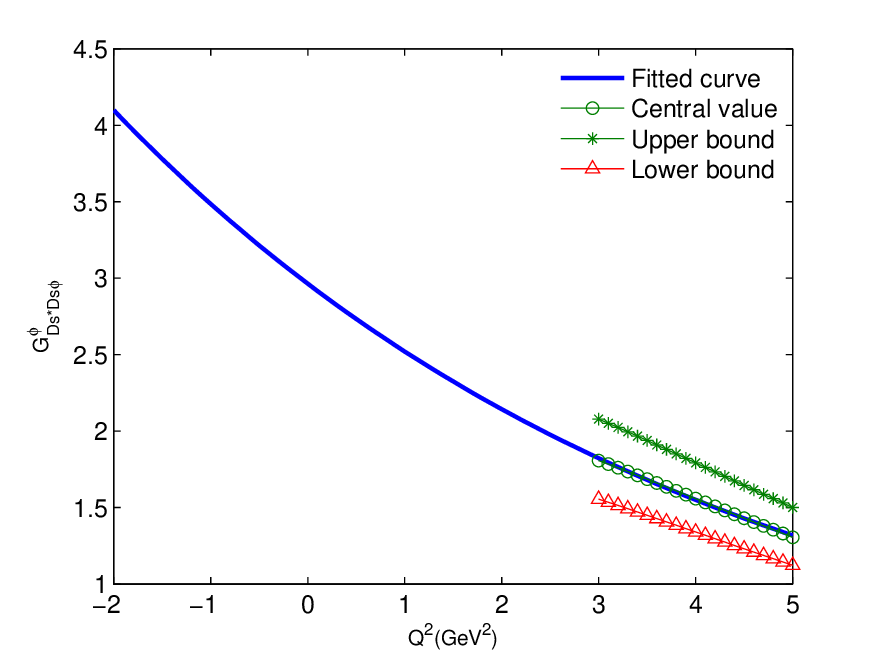}}
\subfigure[]{\includegraphics[height=4.0cm,width=4.6cm]{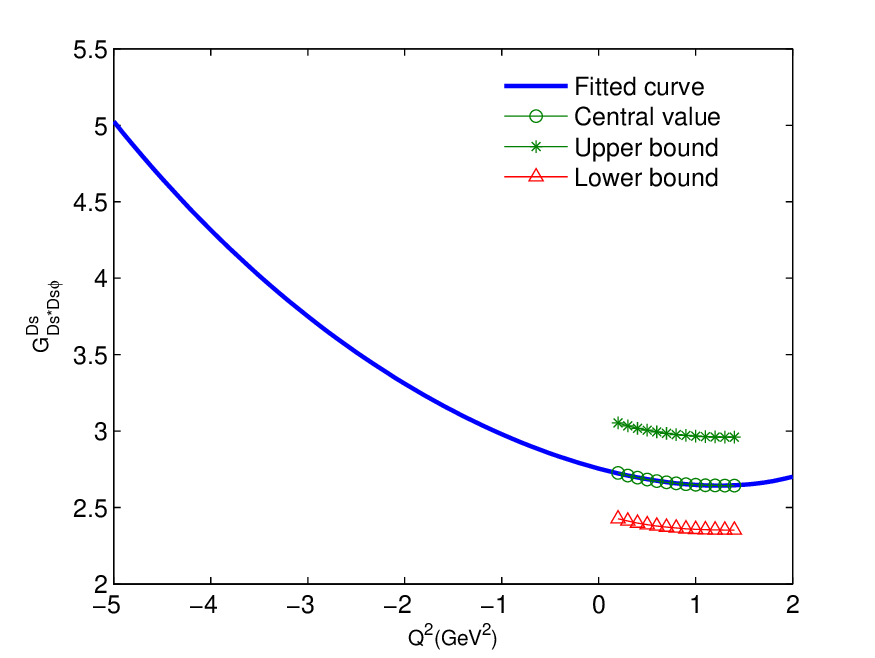}}
\subfigure[]{\includegraphics[height=4.0cm,width=4.6cm]{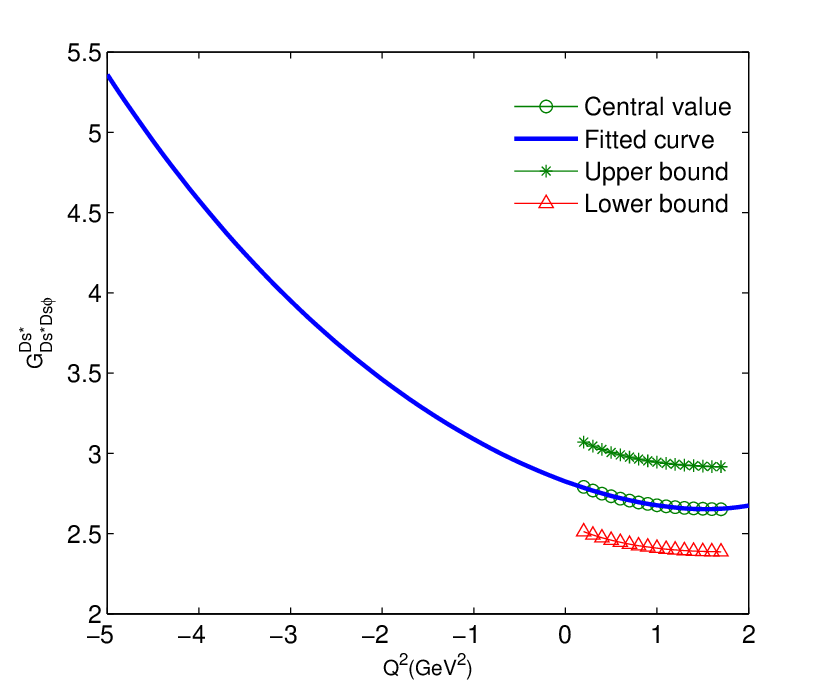}}
\end{center}
\caption{Form factors of the $D_{s}^{*}D_{s}\phi$ vertex for $\phi$
off-shell(a), $D_{s}$ off-shell(b) and $D_{s}^{*}$ off-shell(c).}
\end{figure}

\begin{eqnarray}
&&G^{\phi}_{D_{s}^{*}D_{s}\phi}=Aexp(-BQ^2),\\
&&G^{D_{s}}_{D_{s}^{*}D_{s}\phi}=\frac{C}{1+DQ^2}exp(-EQ^2),\\
&&G^{D^{*}_{s}}_{D_{s}^{*}D_{s}\phi}=\frac{C'}{1+D'Q^2}exp(-E'Q^2),
\end{eqnarray}
where

$A=2.964\pm0.089 GeV^{-1},$ $B=0.1621\pm0.0077GeV^{-2}$,

$C=2.755\pm0.008,$ $D=-0.1944\pm0.0186,$ $E=0.256\pm0.0265,$

$C^{'}=2.825\pm0.012GeV^{-1},$ $D^{'}=-0.1855\pm0.0171GeV^{-2},$
$E'=0.2593\pm0.0257$,

Considering uncertainties of all the input parameters, such as quark
and mesons masses, decay constants and the vlues of different
condensates terms, we plot the upper and lower bounds of the form
factors in Fig.5. We can see that although the uncertainties are
large(about $20\%\sim30\%$ of the central values), the fitted
functions can reproduce the central values of the form factors well.
Thus, it is reliable for us to extrapolate the $Q^2$ to the physical
region $Q^2<0$ for $\phi$, $D_{s}$ and $D_{s}^{*}$ off-shell cases
to obtain the coupling constant for the vertex $D_{s}^{*}D_{s}\phi$.
Substituting $Q^{2}=-m_{\phi}^{2}$, $Q^{2}=-m_{D_{s}}^{2}$ and
$Q^{2}=-m_{D_{s}^{*}}^{2}$ separately in Eqs.$(34)\sim(36)$, we
obtain the values for $G^{\phi}_{D_{s}^{*}D_{s}\phi}$,
$G^{D_{s}}_{D_{s}^{*}D_{s}\phi}$ and
$G^{D_{s}^{*}}_{D_{s}^{*}D_{s}\phi}$:

\begin{eqnarray}
 G^{\phi}_{Ds*Ds\phi}=3.51\pm0.11 GeV^{-1}
\end{eqnarray}
\begin{eqnarray}
 G^{Ds}_{Ds*Ds\phi}=4.24\pm0.47  GeV^{-1}
\end{eqnarray}
\begin{eqnarray}
 G^{Ds*}_{Ds*Ds\phi}=4.61\pm0.51  GeV^{-1}
\end{eqnarray}
Although the values for each off-shell case are different, they are
roughly compatible when the uncertainties are taken into account,
where the uncertainties in Eqs.$(37)\sim(39)$ originate from the
uncertainties of the fitted parameters $\delta A$, $\delta B$,
$\delta C$, $\delta D$, $\delta E$, $\delta C'$, $\delta D'$ and
$\delta E'$. Taking the mean values between the numbers presented
above, we obtain the strong coupling constant for $G_{Ds*Ds\phi}$:
\begin{eqnarray}
G_{Ds*Ds\phi}=4.12\pm0.70 GeV^{-1}
\end{eqnarray}
In Reference\cite{ZGW07}, Z.G.Wang studied the $D^{*}DV$ vertex with
the light-cone QCD sum rules. Most of the results in this work is
analyzed to be much smaller than others\cite{ZGW07}. And the
coupling constant $G_{Ds*Ds\phi}$ is estimated to be about
$0.82\pm0.16GeV^{-1}$ which is also much smaller than our result.
This difference is most probably due to the different input
parameters and the different methods employed. In
Reference\cite{Kohs14}, it is indicated that the value of
$G_{Ds*Ds\phi}$ is $4.07\pm0.71GeV^{-1}$ in the framework of the
three-point QCD sum rules. Besides of the perturbative part, the
contributions of quark-quark, gluon-gluon, and quark-gluon
condensate are considered in this work. It is clearly that our
result is compatible well with that of Referenc\cite{Kohs14}, which
indicates to some extent the reliability of our result.

As to the coupling constant $G_{D_{s}^{*}D_{s}\gamma}$ in the decay
$D_{s}^{*}\rightarrow D_{s}\gamma$ in Eq.(2), we can easily obtain
its value by setting $Q^2=0$ in the analytical function(Eq.(34)):
\begin{eqnarray}
G_{Ds*Ds\gamma}=2.96\pm0.09 GeV^{-1}
\end{eqnarray}
Now, it is time for us to give an analysis of electromagnetic decay
$D_{s}^{*}\rightarrow D_{s}\gamma$. As to its decay width, it can be
written as the following representation:
\begin{eqnarray}
\Gamma =\frac{1}{2J+1}\sum \frac{p}{8\pi M_{i}^{2}}\left\vert
T\right\vert ^{2}
\end{eqnarray}
\begin{eqnarray}
\notag\
p=\frac{\sqrt{%
(M_{i}^{2}-(M_{f}^{2}+m^{2})^{2})(M_{i}^{2}-(M_{f}^{2}-m^{2})^{2})}}{2M_{i}}
\end{eqnarray}
where the $i$ and $f$ denote the initial and final state mesons,
respectively, the $J$ is the total angular momentum of the initial
meson, the $\sum$ denotes the summation of all the polarization
vectors, and the $T$ denotes the scattering amplitudes.

With the Eqs.(33) and (42), the decay width of $D_{s}^{*}\rightarrow
D_{s}\gamma$ can be expressed as:
\begin{eqnarray}
\Gamma =\alpha G_{D_{s}^{*}D_{s}\gamma}^{2}\Big[\frac{f_{\phi }}{m_{\phi }}Q_{s}\Big]^{2}\frac{\left\vert p\right\vert ^{3}}{%
3M_{D_{s}^{\ast }}^{2}}\Big[\frac{M_{D_{s}^{\ast }}^{2}+M_{D_{s}}^{2}}{%
2M_{D_{s}^{\ast }}}\Big]^{2}
\end{eqnarray}
with $Q_{s}=\frac{1}{3}$, $\alpha=\frac{1}{137}$. Considering all
the uncertainties of the input parameters, we finally get the decay
width of the process $D_{s}^{*}\rightarrow D_{s}\gamma$:
\begin{eqnarray}
\notag\ \Gamma=0.59\pm0.15keV
\end{eqnarray}
It is indicated by the Babar collaboration that the decay width of
$\Gamma(D_{s}^{*})<1.9MeV$. And the ratio of the decay channel
$D_{s}^{*}\rightarrow D_{s}\gamma$ is about $94.2\%$ of the total
width. This means that our result is compatible with the
experimental data. Besides, Donald \emph{et al}. predicted the value
of this decay channel is $\Gamma(D_{s}^{*}\rightarrow
D_{s}\gamma)=0.066\pm0.026keV$ in Full Lattice QCD\cite{Dona14},
which is much smaller than our result. Although these results are
all compatible with the experimental data, it needs to be further
testified by more experiments and theoretical calculations because
of this difference.

\begin{large}
\textbf{4 Conclusions}
\end{large}

In this article, we have calculated the form factors
$G^{\phi}_{D_{s}^{*}D_{s}\phi}$, $G^{D_{s}}_{D_{s}^{*}D_{s}\phi}$
and $G^{D_{s}^{*}}_{D_{s}^{*}D_{s}\phi}$ in the space-like regions
with $\phi$, $D_{s}$ and $D_{s}^{*}$ off-shell cases by three
different QCD sum rules. Then we fit the form factors into
analytical functions, extrapolated them into the time-like regions,
and obtained the strong coupling constant $G_{Ds*Ds\phi}$. This
procedure help us to reduce the errors related to the method,
leading to compatible coupling constants, as seen
Eqs.$(37)\sim(39)$. In addition, we also obtained the coupling
constant $G_{Ds*Ds\gamma}$ with the analytical function. With this
coupling constant, we calculated the decay width of the
electromagnetic decay $D_{s}^{*}\rightarrow D_{s}\gamma$ and
compared our result with those of other groups.

\begin{large}
\textbf{Acknowledgment}
\end{large}

This work is supported by National Natural Science Foundation of
China, Grant Number 11375063, the Fundamental Research Funds for the
Central Universities, Grant Number 13QN59,2014ZD42 and the Natural
Science Foundation of GuiZhou Province of China 2013GZ62432.

\begin{large}
\newpage
\textbf{Appendix A: Full expressions of the $<g^{2}G^{2}>$ and
$<f^{3}G^{3}>$ contributions for $\phi$ off-shell case.}
\end{large}

\begin{eqnarray}
\notag\
\Pi_{\mu\nu}^{\left\langle g^{2}G^{2}\right\rangle
(\phi)}&=&\frac
{i\left\langle \frac{g^{2}G^{2}}{4\pi^{2}}\right\rangle }{4\pi^{2}}%
\Bigg \{-m_{c}\Big[\widetilde{I}_{m_{c}m_{s}m_{s}}^{311}+m_{c}^{2}\widetilde{I}%
_{m_{c}m_{s}m_{s}}^{411}+I_{m_{c}m_{s}m_{s}}^{311}+m_{c}^{2}I_{m_{c}m_{s}%
m_{s}}^{411}+N_{m_{c}m_{s}m_{s}}^{311}\\ \notag\
&+&m_{c}^{2}N_{m_{c}m_{s}m_{s}}^{411}\Big]+m_{c}^{2}%
m_{s}\Big[\widetilde{I}_{m_{c}m_{s}m_{s}}^{411}+m_{c}^{2}I_{m_{c}m_{s}m_{s}}%
^{411}\Big]+m_{s}I_{m_{c}m_{s}m_{s}}^{131}\\ \notag\
&+&m_{s}^{2}(m_{s}-m_{c})\Big[I_{m_{c}m_{s}m_{s}%
}^{141}+\widetilde{I}_{m_{c}m_{s}m_{s}}^{141}\Big]-m_{c}m_{s}^{2}N_{m_{c}%
m_{s}m_{s}}^{141}-m_{s}^{2}(m_{c}-m_{s})\Big[I_{m_{c}m_{s}m_{s}}^{114}\\
\notag\
&+&\widetilde{I}_{m_{c}%
m_{s}m_{s}}^{114}\Big]+m_{s}\widetilde{I}_{m_{c}m_{s}m_{s}}^{113}-m_{c}m_{s}%
^{2}N_{m_{c}m_{s}m_{s}}^{114}+\frac{1}{6}\Big[(m_{c}-m_{s})I_{m_{c}m_{s}m_{s}}^{122}\\
\notag\ &+&(m_{c}-m_{s})\widetilde
{I}_{m_{c}m_{s}m_{s}}^{122}+m_{c}N_{m_{c}m_{s}m_{s}}^{122}\Big]+\frac{1}{6}\Big[(m_{c}-m_{s})I_{m_{c}m_{s}m_{s}}^{221}+(m_{c}-m_{s})\widetilde
{I}_{m_{c}m_{s}m_{s}}^{221}\\
\notag\
&+&m_{c}N_{m_{c}m_{s}m_{s}}^{221}\Big]+\frac{1}{6}\Big[3(m_{s}-m_{c})\widetilde{I}_{m_{c}m_{s}m_{s}}^{212}+(m_{s}%
-3m_{c})I_{m_{c}m_{s}m_{s}}^{212}\\
&-&3m_{c}N_{m_{c}m_{s}m_{s}}^{212}\Big]\Bigg \}p_{\alpha }p_{\beta
}^{\prime
}\varepsilon ^{\mu \nu \alpha \beta } \\
\notag\
\Pi _{\mu \nu }^{\left\langle f^{3}G^{3}\right\rangle (\phi )}&=&\frac{%
i\left\langle f^{3}G^{3}\right\rangle }{32\pi ^{4}}\Bigg \{\frac{m_{c}}{6}%
\Big[N_{m_{c}m_{s}m_{s}}^{321}+2m_{c}^{2}N_{m_{c}m_{s}m_{s}}^{421}-3\widetilde{I}%
_{m_{c}m_{s}m_{s}}^{321}-3I_{m_{c}m_{s}m_{s}}^{321}-2m_{c}^{2}\widetilde{I}%
_{m_{c}m_{s}m_{s}}^{421}\\
\notag\ &-&2m_{c}^{2}I_{m_{c}m_{s}m_{s}}^{421}\Big]+\frac{m_{c}}{6}%
\Big[11N_{m_{c}m_{s}m_{s}}^{312}+11m_{c}^{2}N_{m_{c}m_{s}m_{s}}^{412}-2I_{m_{c}m_{s}m_{s}}^{402}+I_{m_{c}m_{s}m_{s}}^{312}\\
\notag\
&+&m_{c}^{2}I_{m_{c}m_{s}m_{s}}^{412}+9\widetilde{I}_{m_{c}m_{s}m_{s}}^{312}+6m_{c}^{2}\widetilde{I}%
_{m_{c}m_{s}m_{s}}^{412}\Big]-\frac{m_{c}}{3}\Big[I_{m_{c}m_{s}m_{s}}^{132}+\widetilde{I}%
_{m_{c}m_{s}m_{s}}^{132}+N_{m_{c}m_{s}m_{s}}^{132}\Big]\\ \notag\
&-&\frac{m_{c}}{6}%
\Big[I_{m_{c}m_{s}m_{s}}^{231}+\widetilde{I}%
_{m_{c}m_{s}m_{s}}^{231}
+N_{m_{c}m_{s}m_{s}}^{231}\Big]-\frac{m_{c}}{6}\Big[I_{m_{c}m_{s}m_{s}}^{123}+\widetilde{I}%
_{m_{c}m_{s}m_{s}}^{123}+N_{m_{c}m_{s}m_{s}}^{123}\Big]\\ \notag\
&+&m_{c}\Big[I_{m_{c}m_{s}m_{s}}^{213}+%
\widetilde{I}_{m_{c}m_{s}m_{s}}^{213}+N_{m_{c}m_{s}m_{s}}^{213}\Big]+\frac{1}{2}\Big[(6m_{s}-m_{c})(\widetilde{I}%
_{m_{c}m_{s}m_{s}}^{411}-I_{m_{c}m_{s}m_{s}}^{411})\\
\notag\
&+&8m_{s}^{2}(m_{s}-m_{c})(%
\widetilde{I}_{m_{c}m_{s}m_{s}}^{511}-I_{m_{c}m_{s}m_{s}}^{511})+(m_{s}-m_{c})(I_{m_{c}m_{s}m_{s}}^{411}+8m_{s}^{2}I_{m_{c}m_{s}m_{s}}^{511})\\
\notag\
&-&6m_{s}N_{m_{c}m_{s}m_{s}}^{411}-8m_{s}^{3}N_{m_{c}m_{s}m_{s}}^{511}\Big]+\frac{1}{2}%
\Big[(m_{s}-6m_{c})I_{m_{c}m_{s}m_{s}}^{411}+8m_{c}^{2}(m_{s}-m_{c})I_{m_{c}m_{s}m_{s}}^{511}\\
\notag\
&-&(m_{s}-6m_{c})%
\widetilde{I}_{m_{c}m_{s}m_{s}}^{411}-8m_{c}^{2}(m_{s}-m_{c})\widetilde{I}%
_{m_{c}m_{s}m_{s}}^{511} -6m_{c}N_{m_{c}m_{s}m_{s}}^{411}
-8m_{c}^{3}N_{m_{c}m_{s}m_{s}}^{511}\Big]\\
\notag\
&+&\frac{%
1}{2}\Big[(m_{s}-m_{c})(\widetilde{I}%
_{m_{c}m_{s}m_{s}}^{141}-I_{m_{c}m_{s}m_{s}}^{141}+9m_{s}^{2}\widetilde{I}%
_{m_{c}m_{s}m_{s}}^{151}-9m_{s}^{2}I_{m_{c}m_{s}m_{s}}^{151})\\
\notag\
&+&(m_{s}-m_{c})(I_{m_{c}m_{s}m_{s}}^{141}+9m_{s}^{2}I_{m_{c}m_{s}m_{s}}^{151})+5m_{s}(I_{m_{c}m_{s}m_{s}}^{141}-m_{s}^{2}I_{m_{c}m_{s}m_{s}}^{151})\\
\notag\
&+&m_{s}(N_{m_{c}m_{s}m_{s}}^{141}+9m_{s}^{2}N_{m_{c}m_{s}m_{s}}^{151})\Big]-\frac{m_{c}}{6}\Big[I_{m_{c}m_{s}m_{s}}^{222}+\widetilde{I}%
_{m_{c}m_{s}m_{s}}^{222}+N_{m_{c}m_{s}m_{s}}^{222}\Big]\Bigg
\}p_{\alpha
}p_{\beta }^{\prime }\varepsilon ^{\mu \nu \alpha \beta } \\
\end{eqnarray}
\newpage
\begin{large}
\textbf{Appendix B: Full expressions about the condensate terms for
$D_{s}$ off-shell case.}
\end{large}

\begin{eqnarray}
\notag\
\Pi _{\mu \nu }^{\left\langle \overline{s}s\right\rangle (D_{s})}&=&\frac{%
\left\langle \overline{s}s\right\rangle }{2}\Bigg
\{-\frac{m_{s}^{2}}{(p^{\prime
2}-m_{c}^{2})(p^{2}-m_{s}^{2})^{2}}+\frac{m_{c}m_{s}}{(p^{\prime
2}-m_{c}^{2})^{2}(p^{2}-m_{s}^{2})}+\frac{2}{(p^{\prime
2}-m_{c}^{2})(p^{2}-m_{s}^{2})}\Bigg \}p_{\alpha }p_{\beta }^{\prime
}\varepsilon ^{\mu \nu \alpha \beta }\\
\\
\notag\ \Pi _{\mu \nu }^{\left\langle \overline{s}g\sigma
.Gs\right\rangle
(D_{s})}&=&\left\langle \overline{s}g\sigma .Gs\right\rangle \Bigg \{\frac{1}{12}\Big[%
\frac{1}{(p^{\prime
2}-m_{c}^{2})^{2}(p^{2}-m_{s}^{2})}+\frac{1}{(p^{\prime
2}-m_{c}^{2})(p^{2}-m_{s}^{2})^{2}}\Big]\\ \notag\
&-&\frac{1}{96}\Big[-2m_{s}^{2}\big(\frac{12}{%
(p^{\prime 2}-m_{c}^{2})(p^{2}-m_{s}^{2})^{3}}+\frac{24m_{s}^{2}}{(p^{\prime 2}-m_{c}^{2})(p^{2}-m_{s}^{2})^{4}}+\frac{2}{%
(p^{\prime 2}-m_{c}^{2})^{2}(p^{2}-m_{s}^{2})^{2}}\\ \notag\
&+&\frac{8m_{c}^{2}}{%
(p^{\prime 2}-m_{c}^{2})^{3}(p^{2}-m_{s}^{2})^{2}}\big)+2m_{c}m_{s}\big(\frac{12}{%
(p^{\prime
2}-m_{c}^{2})^{3}(p^{2}-m_{s}^{2})}+\frac{24m_{c}^{2}}{(p^{\prime
2}-m_{c}^{2})^{4}(p^{2}-m_{s}^{2})}\\ \notag\
&+&\frac{2}{(p^{\prime 2}-m_{c}^{2})^{2}(p^{2}-m_{s}^{2})^{2}}+\frac{%
8m_{s}^{2}}{(p^{\prime 2}-m_{c}^{2})^{2}(p^{2}-m_{s}^{2})^{3}}\big)+6\big(\frac{2}{%
(p^{\prime 2}-m_{c}^{2})(p^{2}-m_{s}^{2})^{2}}\\
\notag\ &+&\frac{8m_{s}^{2}}{(p^{\prime
2}-m_{c}^{2})(p^{2}-m_{s}^{2})^{3}}+\frac{2}{(p^{\prime
2}-m_{c}^{2})^{2}(p^{2}-m_{s}^{2})}+\frac{8m_{c}^{2}}{(p^{\prime
2}-m_{c}^{2})^{3}(p^{2}-m_{s}^{2})}\big)\Big]\Bigg \}p_{\alpha
}p_{\beta }^{\prime
}\varepsilon ^{\mu \nu \alpha \beta}\\
\\
\notag\ \Pi _{\mu \nu }^{\left\langle \overline{s}s\right\rangle
^{2}(D_{s})}&=&\Bigg\{ \frac{2\left\langle \overline{s}s\right\rangle ^{2}m_{c}}{%
27(p^{2}-m_{s}^{2})(p^{\prime 2}-m_{c}^{2})^{3}}-\frac{\left\langle
\overline{s}s\right\rangle ^{2}}{162(p^{2}-m_{s}^{2})(p^{\prime 2}-m_{c}^{2})%
}\Big [\frac{3m_{s}}{(p^{2}-m_{s}^{2})^{2}}+\frac{2m_{s}^{3}}{%
(p^{2}-m_{s}^{2})^{3}}\\ \notag\
&-&\frac{m_{c}}{(p^{\prime 2}-m_{c}^{2})(p^{2}-m_{s}^{2})%
}+\frac{2m_{c}m_{s}^{2}}{(p^{\prime
2}-m_{c}^{2})(p^{2}-m_{s}^{2})^{2}}+\frac{m_{s}}{(p^{\prime
2}-m_{c}^{2})(p^{2}-m_{s}^{2})}\\&+&\frac{2m_{s}m_{c}^{2}}{%
(p^{\prime2}-m_{c}^{2})^{2}(p^{2}-m_{s}^{2})}-\frac{3m_{c}}{(p^{\prime
2}-m_{c}^{2})^{2}}-\frac{2m_{c}^{3}}{(p^{\prime
2}-m_{c}^{2})^{3}}\Big ]\Bigg \} p_{\alpha }p_{\beta }^{\prime
}\varepsilon ^{\mu
\nu \alpha \beta}\\
\notag\
\Pi _{\mu \nu }^{\left\langle \overline{s}sGG\right\rangle (D_{s})}&=&\frac{%
\left\langle g^{2}\overline{s}sGG\right\rangle }{72}\Bigg \{-\frac{3m_{s}^{2}}{%
(p^{\prime 2}-m_{c}^{2})^{2}(p^{2}-m_{s}^{2})^{3}}+\frac{3m_{c}m_{s}}{%
(p^{\prime
2}-m_{c}^{2})^{3}(p^{2}-m_{s}^{2})^{2}}+\frac{1}{(p^{\prime
2}-m_{c}^{2})^{2}(p^{2}-m_{s}^{2})^{2}}\\
\notag\ &+&\frac{2}{(p^{\prime
2}-m_{c}^{2})^{2}(p^{2}-m_{s}^{2})^{2}}+\frac{4m_{s}^{2}}{(p^{\prime
2}-m_{c}^{2})^{2}(p^{2}-m_{s}^{2})^{3}}+\frac{4m_{c}^{2}}{(p^{\prime
2}-m_{c}^{2})^{3}(p^{2}-m_{s}^{2})^{2}}\\ \notag\
&+&\frac{4m_{s}^{2}m_{c}^{2}}{(p^{\prime
2}-m_{c}^{2})^{3}(p^{2}-m_{s}^{2})^{3}}+\frac{3}{(p^{\prime
2}-m_{c}^{2})(p^{2}-m_{s}^{2})^{3}}+\frac{12m_{s}^{2}}{(p^{\prime
2}-m_{c}^{2})(p^{2}-m_{s}^{2})^{4}}\\
\notag\ &+&\frac{4m_{s}^{4}}{(p^{\prime
2}-m_{c}^{2})(p^{2}-m_{s}^{2})^{5}}+\frac{3}{(p^{\prime
2}-m_{c}^{2})^{3}(p^{2}-m_{s}^{2})}+\frac{12m_{c}^{2}}{(p^{\prime
2}-m_{c}^{2})^{4}(p^{2}-m_{s}^{2})}\\
\notag\ &+& \frac{4m_{c}^{4}}{(p^{\prime
2}-m_{c}^{2})^{5}(p^{2}-m_{s}^{2})}+\frac{m_{s}^{2}m_{c}m_{s}}{2(p^{\prime
2}-m_{c}^{2})^{2}(p^{2}-m_{s}^{2})^{4}}-\frac{m_{s}^{4}}{2(p^{\prime
2}-m_{c}^{2})(p^{2}-m_{s}^{2})^{5}}\\
\notag\ &-&\frac{m_{s}^{2}}{2(p^{\prime
2}-m_{c}^{2})(p^{2}-m_{s}^{2})^{4}}+\frac{m_{c}^{2}m_{c}m_{s}}{2(p^{\prime
2}-m_{c}^{2})^{5}(p^{2}-m_{s}^{2})}-\frac{m_{c}^{2}m_{s}^{2}}{2(p^{\prime
2}-m_{c}^{2})^{4}(p^{2}-m_{s}^{2})^{2}}\\
 &+&\frac{3m_{c}m_{s}}{2(p^{\prime
2}-m_{c}^{2})^{4}(p^{2}-m_{s}^{2})}+\frac{2m_{c}^{2}}{2(p^{\prime
2}-m_{c}^{2})^{4}(p^{2}-m_{s}^{2})}\Bigg \}p_{\alpha }p_{\beta
}^{\prime }\varepsilon ^{\mu \nu \alpha \beta}
\end{eqnarray}

\begin{eqnarray}
\notag\
\Pi _{\mu \nu }^{\left\langle g^{2}G^{2}\right\rangle (D_{s})}&=&\frac{%
i\left\langle \frac{g^{2}G^{2}}{4\pi ^{2}}\right\rangle }{4\pi ^{2}}%
\Bigg \{-m_{s}\Big[(m_{s}^{2}-m_{s}m_{c})\widetilde{I}_{m_{s}m_{s}m_{c}}^{411}+%
\widetilde{I}%
_{m_{s}m_{s}m_{c}}^{311}+N_{m_{s}m_{s}m_{c}}^{311}+m_{s}^{2}N_{m_{s}m_{s}m_{c}}^{411}\\
\notag\ &+&I_{m_{s}m_{s}m_{c}}^{311}\Big]+m_{s}\Big[m_{s}(m_{c}-m_{s})\widetilde{I}%
_{m_{s}m_{s}m_{c}}^{141}-m_{s}^{2}N_{m_{s}m_{s}m_{c}}^{141}+I_{m_{s}m_{s}m_{c}}^{131}\Big]+m_{c}\Big[%
\widetilde{I}_{m_{s}m_{s}m_{c}}^{113}\\ \notag\
&+&(m_{c}^{2}-m_{s}m_{c})\widetilde{I}%
_{m_{s}m_{s}m_{c}}^{114}-m_{s}m_{c}N_{m_{s}m_{s}m_{c}}^{114}\Big]+\frac{1}{6}\Big[(3m_{c}-m_{s})\widetilde{I}%
_{m_{s}m_{s}m_{c}}^{122}+2m_{s}I_{m_{s}m_{s}m_{c}}^{122}\\ \notag\
&-&m_{s}N_{m_{s}m_{s}m_{c}}^{122}\Big]+%
\frac{1}{6}\Big[(m_{s}-m_{c})\widetilde{I}%
_{m_{s}m_{s}m_{c}}^{221}+m_{s}N_{m_{s}m_{s}m_{c}}^{221}\Big]+\frac{1}{6}\Big[(m_{s}-m_{c})\widetilde{I}%
_{m_{s}m_{s}m_{c}}^{212}\\
&+&m_{s}N_{m_{s}m_{s}m_{c}}^{212}\Big]\Bigg \}p_{\alpha }p_{\beta
}^{\prime
}\varepsilon ^{\mu \nu \alpha \beta}\\
\notag\
\Pi _{\mu \nu }^{\left\langle f^{3}G^{3}\right\rangle (D_{s})}&=&\frac{%
i\left\langle f^{3}G^{3}\right\rangle }{32\pi ^{4}}\Bigg \{\frac{m_{c}}{6}%
\widetilde{I}_{m_{s}m_{s}m_{c}}^{321}+\frac{m_{c}}{3}\widetilde{I}%
_{m_{s}m_{s}m_{c}}^{312}-m_{c}\widetilde{I}_{m_{s}m_{s}m_{c}}^{132}+\frac{%
m_{c}}{6}\widetilde{I}_{m_{s}m_{s}m_{c}}^{231}\\ \notag\
&+&\frac{m_{c}}{6}\Big[-5\widetilde{I}%
_{m_{s}m_{s}m_{c}}^{123}+2I_{m_{s}m_{s}m_{c}}^{123}+2I_{m_{s}m_{s}m_{c}}^{114}-2m_{c}^{2}%
\widetilde{I}_{m_{s}m_{s}m_{c}}^{124}+2(m_{s}^{2}+m_{c}^{2}\\
\notag\
&+&q\symbol{94}%
2)I_{m_{s}m_{s}m_{c}}^{124}\Big]
+\frac{m_{c}}{6}%
\Big[4N_{m_{s}m_{s}m_{c}}^{213}+4m_{c}^{2}N_{m_{s}m_{s}m_{c}}^{214}+\widetilde{I}%
_{m_{s}m_{s}m_{c}}^{213}+2\widetilde{I}_{m_{s}m_{s}m_{c}}^{114}\Big]\\
\notag\ &+&m_{c}\Big[3\widetilde{I}_{m_{s}m_{s}m_{c}}^{114}+4m_{c}^{2}\widetilde{I}%
_{m_{s}m_{s}m_{c}}^{115}\Big]+\frac{m_{c}}{2}\widetilde{I}%
_{m_{s}m_{s}m_{c}}^{411}+\frac{1}{2}\Big[m_{c}\widetilde{I}_{m_{s}m_{s}m_{c}}^{141}+8m_{s}^{2}m_{c}%
\widetilde{I}%
_{m_{s}m_{s}m_{c}}^{151}\\ \notag\
&-&6m_{s}I_{m_{s}m_{s}m_{c}}^{141}-8m_{s}^{3}I_{m_{s}m_{s}m_{c}}^{151}\Big]+%
\frac{1}{6}\Big[(m_{c}-m_{s})\widetilde{I}%
_{m_{s}m_{s}m_{c}}^{222}-m_{s}N_{m_{s}m_{s}m_{c}}^{222}\Big]\Bigg
\}p_{\alpha
}p_{\beta }^{\prime }\varepsilon ^{\mu \nu \alpha \beta} \\
\end{eqnarray}

\newpage
\begin{large}
\textbf{Appendix C: Full expressions about the condensate terms for
$D_{s}^{*}$ off-shell case.}
\end{large}

\begin{eqnarray}
\notag\ \Pi _{\mu \nu }^{\left\langle \overline{s}s\right\rangle
(D_{s}^{\ast })}&=&
\frac{\left\langle \overline{s}s\right\rangle }{2}\Bigg \{\frac{m_{s}^{2}}{%
(p^{\prime
2}-m_{s}^{2})^{2}(p^{2}-m_{c}^{2})}-\frac{m_{c}m_{s}}{(p^{\prime
2}-m_{s}^{2})(p^{2}-m_{c}^{2})^{2}}+\frac{2}{(p^{\prime
2}-m_{s}^{2})(p^{2}-m_{c}^{2})}\Bigg\}p_{\alpha }p_{\beta }^{\prime
}\varepsilon ^{\mu \nu \alpha \beta }\\
\\
\notag\ \Pi _{\mu \nu }^{\left\langle \overline{s}s\right\rangle
^{2}(D_{s}^{\ast })}&=&\Bigg\{\frac{2\left\langle \overline{s}s\right\rangle ^{2}}{%
27(p^{2}-m_{c}^{2})(p^{\prime 2}-m_{s}^{2})}\Big[\frac{m_{c}}{%
(p^{2}-m_{c}^{2})^{2}}+\frac{m_{s}}{(p^{\prime 2}-m_{s}^{2})^{2}}\Big]-\frac{%
\left\langle \overline{s}s\right\rangle
^{2}}{162(p^{2}-m_{c}^{2})(p^{\prime 2}-m_{s}^{2})}\\ \notag\
&\times& \Big[\frac{3m_{c}}{(p^{2}-m_{c}^{2})^{2}}+\frac{2m_{c}^{3}}{%
(p^{2}-m_{c}^{2})^{3}}-\frac{m_{s}}{(p^{\prime 2}-m_{s}^{2})(p^{2}-m_{c}^{2})%
}-\frac{2m_{s}m_{c}^{2}}{(p^{\prime
2}-m_{s}^{2})(p^{2}-m_{c}^{2})^{2}}\\ \notag\
&+&\frac{m_{c}}{(p^{\prime 2}-m_{s}^{2})(p^{2}-m_{c}^{2})}+\frac{%
2m_{c}m_{s}^{2}}{(p^{\prime 2}-m_{s}^{2})^{2}(p^{2}-m_{c}^{2})}-\frac{3m_{s}%
}{(p^{\prime 2}-m_{s}^{2})^{2}}-\frac{2m_{s}^{3}}{(p^{\prime
2}-m_{s}^{2})^{3}}\Big]\Bigg\}p_{\alpha }p_{\beta }^{\prime
}\varepsilon ^{\mu \nu \alpha \beta }\\
\\
\notag\ \Pi _{\mu \nu }^{\left\langle \overline{s}g\sigma
.Gs\right\rangle
(D_{s}^{\ast })}&=&\left\langle \overline{s}g\sigma .Gs\right\rangle \Bigg\{\frac{1%
}{12}\Big[\frac{1}{(p^{\prime 2}-m_{s}^{2})^{2}(p^{2}-m_{c}^{2})}-\frac{3}{%
(p^{\prime 2}-m_{s}^{2})(p^{2}-m_{c}^{2})^{2}}\Big]\\ \notag\
&-&\frac{1}{24}\Big[-m_{s}^{2}\big(%
\frac{3}{(p^{\prime 2}-m_{s}^{2})^{3}(p^{2}-m_{c}^{2})}
+\frac{2m_{s}^{2}}{(p^{\prime 2}-m_{s}^{2})^{4}(p^{2}-m_{c}^{2})}+\frac{1}{%
(p^{\prime 2}-m_{s}^{2})^{2}(p^{2}-m_{c}^{2})^{2}}\\ \notag\
&+&\frac{2m_{c}^{2}}{%
(p^{\prime 2}-m_{s}^{2})^{2}(p^{2}-m_{c}^{2})^{3}}\big)+m_{c}m_{s}\big(\frac{3}{%
(p^{\prime
2}-m_{s}^{2})(p^{2}-m_{c}^{2})^{3}}+\frac{2m_{c}^{2}}{(p^{\prime
2}-m_{s}^{2})(p^{2}-m_{c}^{2})^{4}}\\ \notag\
&+&\frac{1}{(p^{\prime 2}-m_{s}^{2})^{2}(p^{2}-m_{c}^{2})^{2}}+\frac{%
2m_{s}^{2}}{(p^{\prime 2}-m_{s}^{2})^{3}(p^{2}-m_{c}^{2})^{2}}\big)+3\big(\frac{1}{%
(p^{\prime 2}-m_{s}^{2})(p^{2}-m_{c}^{2})^{2}}\\ \notag\
&+&\frac{2m_{c}^{2}}{(p^{\prime
2}-m_{s}^{2})(p^{2}-m_{c}^{2})^{3}}+\frac{1}{(p^{\prime
2}-m_{s}^{2})^{2}(p^{2}-m_{c}^{2})}+\frac{2m_{s}^{2}}{(p^{\prime
2}-m_{s}^{2})^{3}(p^{2}-m_{c}^{2})}\big)\Big]\Bigg\}p_{\alpha
}p_{\beta }^{\prime
}\varepsilon ^{\mu \nu \alpha \beta }\\
\end{eqnarray}

\begin{eqnarray}
\notag\
\Pi _{\mu \nu }^{\left\langle \overline{s}sGG\right\rangle
(D_{s}^{*})}&=&\frac{\left\langle g^{2}\overline{s}sGG\right\rangle }{72}\Bigg\{-\frac{%
m_{s}^{2}}{2(p^{\prime 2}-m_{s}^{2})^{3}(p^{2}-m_{c}^{2})^{2}}+\frac{%
m_{c}m_{s}}{2(p^{\prime 2}-m_{s}^{2})^{2}(p^{2}-m_{c}^{2})^{3}}\\
\notag\ &+&\frac{1}{(p^{\prime
2}-m_{s}^{2})^{2}(p^{2}-m_{c}^{2})^{2}}+\frac{2}{(p^{\prime
2}-m_{s}^{2})^{2}(p^{2}-m_{c}^{2})^{2}}+\frac{4m_{s}^{2}}{(p^{\prime
2}-m_{s}^{2})^{2}(p^{2}-m_{c}^{2})^{3}}\\
\notag\ &+&\frac{4m_{c}^{2}}{(p^{\prime
2}-m_{s}^{2})^{3}(p^{2}-m_{c}^{2})^{2}}+\frac{4m_{s}^{2}m_{c}^{2}}{(p^{\prime
2}-m_{s}^{2})^{3}(p^{2}-m_{c}^{2})^{3}}+\frac{3}{(p^{\prime
2}-m_{s}^{2})(p^{2}-m_{c}^{2})^{3}}\\
\notag\ &+&\frac{12m_{s}^{2}}{(p^{\prime
2}-m_{s}^{2})(p^{2}-m_{c}^{2})^{4}}+\frac{4m_{s}^{4}}{(p^{\prime
2}-m_{s}^{2})(p^{2}-m_{c}^{2})^{5}}+\frac{3}{(p^{\prime
2}-m_{s}^{2})^{3}(p^{2}-m_{c}^{2})}\\
\notag\
&+&\frac{12m_{c}^{2}}{(p^{\prime
2}-m_{s}^{2})^{4}(p^{2}-m_{c}^{2})}+\frac{4m_{c}^{4}}{(p^{\prime
2}-m_{s}^{2})^{5}(p^{2}-m_{c}^{2})}\frac{m_{c}^{2}}{(p^{\prime 2}-m_{s}^{2})(p^{2}-m_{c}^{2})^{4}}\\ \notag\
&-&\frac{%
3m_{c}m_{s}}{2(p^{\prime 2}-m_{s}^{2})(p^{2}-m_{c}^{2})^{4}}-\frac{%
m_{s}m_{c}^{3}}{2(p^{\prime 2}-m_{s}^{2})(p^{2}-m_{c}^{2})^{5}}+\frac{%
m_{s}^{2}m_{c}^{2}}{2(p^{\prime 2}-m_{s}^{2})^{2}(p^{2}-m_{c}^{2})^{4}}\\ \notag\
&-&\frac{m_{c}m_{s}^{3}}{2(p^{\prime 2}-m_{s}^{2})^{4}(p^{2}-m_{c}^{2})^{2}}+%
\frac{m_{s}^{4}}{2(p^{\prime 2}-m_{s}^{2})^{5}(p^{2}-m_{c}^{2})}+\frac{%
5m_{s}^{2}}{2(p^{\prime
2}-m_{s}^{2})^{4}(p^{2}-m_{c}^{2})}\Bigg\}p_{\alpha }p_{\beta
}^{\prime
}\varepsilon ^{\mu \nu \alpha \beta} \\
\\ \notag\
\Pi _{\mu \nu }^{\left\langle g^{2}G^{2}\right\rangle
(D_{s}^{\ast })}&=&
\frac{i\left\langle \frac{g^{2}G^{2}}{4\pi^{2}}\right\rangle }{4\pi ^{2}}%
\Bigg\{-m_{s}\Big[(m_{s}^{2}-m_{s}m_{c})I_{m_{s}m_{c}m_{s}}^{411}+I_{m_{s}m_{c}m_{s}}^{311}+N_{m_{s}m_{c}m_{s}}^{311}+m_{s}^{2}N_{m_{s}m_{c}m_{s}}^{411}\\
\notag\ &+&\widetilde{I}_{m_{s}m_{c}m_{s}}^{311}\Big]
+m_{c}\Big[m_{c}(m_{c}-m_{s})I_{m_{s}m_{c}m_{s}}^{141}-m_{s}m_{c}N_{m_{s}m_{c}m_{s}}^{141}+I_{m_{s}m_{c}m_{s}}^{131}\Big]+m_{s}\Big[%
\widetilde{I}%
_{m_{s}m_{c}m_{s}}^{113}\\
\notag\
&+&m_{s}(m_{c}-m_{s})I_{m_{s}m_{c}m_{s}}^{114}-m_{s}^{2}N_{m_{s}m_{c}m_{s}}^{114}\Big]
+\frac{1}{6}%
\Big[(m_{s}-m_{c})I_{m_{s}m_{c}m_{s}}^{122}+m_{s}N_{m_{s}m_{c}m_{s}}^{122}\Big]\\
\notag\
&+&\frac{1}{6}\Big[3(m_{c}-m_{s})I_{m_{s}m_{c}m_{s}}^{221}-2m_{s}\widetilde{I}%
_{m_{s}m_{c}m_{s}}^{221}-3m_{s}N_{m_{s}m_{c}m_{s}}^{221}\Big]\\
&+&\frac{1}{6}\Big[(m_{s}-m_{c})I_{m_{s}m_{c}m_{s}}^{212}+m_{s}N_{m_{s}m_{c}m_{s}}^{212}\Big]\Bigg\}p_{\alpha
}p_{\beta }^{\prime }\varepsilon ^{\mu \nu \alpha \beta} \\ \notag\
\Pi _{\mu \nu }^{\left\langle f^{3}G^{3}\right\rangle
(D_{s}^{\ast })}&=&\frac{i\left\langle f^{3}G^{3}\right\rangle m_{c}}{32\pi ^{4}}%
\Bigg\{-I_{m_{s}m_{c}m_{s}}^{321}+\frac{1}{6}I_{m_{s}m_{c}m_{s}}^{312}+\frac{1}{2}%
I_{m_{s}m_{c}m_{s}}^{132}+\frac{1}{3}\Big[ \widetilde{I}%
_{m_{s}m_{c}m_{s}}^{231}+I_{m_{s}m_{c}m_{s}}^{231}\Big]\\ \notag\
&+&\frac{1}{3}I_{m_{s}m_{c}m_{s}}^{123}+\frac{1}{6}I_{m_{s}m_{c}m_{s}}^{213}+%
\frac{1}{2}I_{m_{s}m_{c}m_{s}}^{114}+\frac{1}{2}I_{m_{s}m_{c}m_{s}}^{411}+3I_{m_{s}m_{c}m_{s}}^{141}-2m_{c}^{2}I_{m_{s}m_{c}m_{s}}^{151}\\
&+&\frac{1}{6}I_{m_{s}m_{c}m_{s}}^{222}\Bigg\}p_{\alpha }p_{\beta
}^{\prime }\varepsilon ^{\mu \nu \alpha \beta}
\end{eqnarray}

\end{document}